\documentclass[jimaging,article,accept,moreauthors,pdftex]{Definitions/mdpi} 

\firstpage{1} 
\pubvolume{1}
\issuenum{1}
\articlenumber{0}
\pubyear{2021}
\copyrightyear{2021}
\externaleditor{Academic Editor: Fabiana Zama; Elena Loli Piccolomini} 
\datereceived{11 August 2021} 
\dateaccepted{10 October 2021} 
\datepublished{} 
\hreflink{https://doi.org/} 

\usepackage{framed}
\usepackage{latexsym}

\usepackage{verbatimbox}
\usetikzlibrary{shapes,arrows}
\usetikzlibrary{fit,backgrounds,positioning}
\usepackage{graphbox}
\usepackage{amsfonts}
\usepackage{amssymb}
\usepackage{amstext}
\usepackage{amsopn}
\usepackage{algorithm}
\usepackage{algpseudocode}
\usepackage{epstopdf}
\usepackage{etoolbox}\AtBeginEnvironment{algorithmic}{\footnotesize}

\Title{Fast Fiber Orientation Estimation in Diffusion MRI from kq-Space Sampling and Anatomical Priors}

\TitleCitation{Fast Fiber Orientation Estimation in Diffusion MRI from kq-Space Sampling and Anatomical Priors}

\Author{Marica Pesce $^{1,}$*, Audrey Repetti $^{1,2,}$*, Anna
 Aur\'ia $^{3}$, Alessandro Daducci $^{4}$, Jean-Philippe Thiran $^{3,5}$ \mbox{and Yves Wiaux $^{1,}$*}}

\AuthorNames{Marica Pesce, Audrey Repetti, Anna Aur\'ia, Alessandro Daducci, Jean-Philippe Thiran and Yves Wiaux}

\AuthorCitation{Pesce, M.; Repetti, A.; Aur\'ia, A.; Daducci, A.; Thiran, J.-P.; Wiaux, Y.}

\address{%
$^{1}$ \quad Institute of Sensors Signals and Systems (ISSS), Heriot-Watt University, Edinburgh EH14~4AS, UK\\
$^{2}$ \quad Department of Actuarial Mathematics \& Statistics (AMS), Heriot-Watt University, Edinburgh EH14~4AS, UK\\
$^{3}$ \quad Signal Processing Lab (LTS5), \'Ecole Polytechnique F\'ed\'erale de Lausanne, CH-1015 Lausanne, Switzerland; anna.auria@gmail.com (A.A.); jean-philippe.thiran@epfl.ch (J.-P.T.)\\ 
$^{4}$ \quad Department of Computer Science, University of Verona, 37134 Verona, Italy; alessandro.daducci@univr.it\\ 
$^{5}$ \quad Department of Radiology, Lausanne University Hospital (CHUV) and University of Lausanne, 1011 Lausanne, Switzerland}

\corres{Correspondence: pesce.marica@gmail.com (M.P.); a.repetti@hw.ac.uk (A.R.); y.wiaux@hw.ac.uk (Y.W.)}

\abstract{High spatio-angular resolution diffusion MRI (dMRI) has been shown to provide accurate identification of complex neuronal fiber configurations, albeit, at the cost of long acquisition times. We propose a method to recover intra-voxel fiber configurations at high spatio-angular resolution relying on a 3D kq-space under-sampling scheme to enable accelerated acquisitions. The inverse problem for the reconstruction of the fiber orientation distribution (FOD) is regularized by a \textit{structured sparsity} prior promoting simultaneously voxel-wise sparsity and spatial smoothness of fiber orientation.
Prior knowledge of the spatial distribution of white matter, gray matter, and cerebrospinal fluid is also leveraged. A minimization problem is formulated and solved via a stochastic forward--backward algorithm. Simulations and real data analysis suggest that accurate FOD mapping can be achieved from severe kq-space under-sampling regimes potentially enabling high spatio-angular resolution dMRI in the clinical setting.}


\begin{document}


\section{Introduction}
\label{Sec:Intro}

Diffusion Magnetic Resonance Imaging (dMRI) is a unique non-invasive technique to infer the microscopic architecture of tissues in vivo.
In recent years, dMRI has gained a lot of attention in neuroscience since it enables the mapping of the white matter fiber paths, revealing the existing connections between different brain areas \cite{jara2013theory,Bihan2003, Sporns2005}. In clinics, dMRI has shown to provide insights into many neurodegenerative diseases, such as Schizophrenia and Alzheimer's disease \cite{Zhang2009, Park2004}. 
{Diffusion MRI enables to map the restricted diffusion of the water molecules comprising the white matter tissue. The information captured by dMRI is then processed in order to infer the connectivity and the integrity of the white matter pathways.}

A typical approach to trace the complex pathways of the white matter fiber bundles {from dMRI signals} relies on piecing together local fiber orientation information. Such local information is obtained by processing a multitude of MRI signals generated by applying various diffusion gradients during the acquisition. Each diffusion-weighted (DW) signal is sensitive to diffusion along a specific direction and at a specific intensity, identified by a so-called q-space point defined by the diffusion gradient applied \cite{Callaghan2000}.

Diffusion Tensor Imaging (DTI) \cite{Basser1994} is the most widely-used technique in clinics for the intra-voxel fiber orientation estimation since it requires less than $7$ diffusion volumes, i.e., $7$ q-points.
{DTI relies on a Gaussian description of the water molecules diffusion phenomenon, which enables the presence of only one preferred diffusion direction. 
For this reason, the presence of multiple fiber populations cannot be distinguished within the same voxel by using DTI, making this technique inappropriate for complex fiber architecture estimation.}

High Angular Resolution Diffusion Imaging (HARDI) methods \cite{Tuch2002a} have shown to provide more accurate fiber representations for complex fiber arrangements, by probing the observed brain with a large number of diffusion gradients.
For instance, Diffusion Spectrum Imaging \cite{Wedeen2005} considers $515$ diffusion gradients allocated on a Cartesian grid, while q-ball approaches \cite{Tuch2004a} use between $30$ and $100$ diffusion gradients distributed over a single shell.
However, acquisitions characterized by large numbers of diffusion gradients result in long scan times, which are unfeasible in clinical applications. 
This limitation becomes even more obvious when the brain microstructure is investigated, encompassing not only fiber orientation, but also other microstructure parameters (e.g., diameter, etc.). 

To this purpose, more elaborate acquisition protocols are used, requiring more diffusion gradients, typically distributed over multiple shells in q-space \cite{Zhang2011, Zhang2012}.
{Consequently, numerous HARDI approaches have been proposed in order to optimize the number and the spatial distribution of the q-space points to be considered for an accurate description on the white matter fiber arrangement \cite{Daducci2014,Ning2015}}.
More specifically, depending on the micro-structure parameters of interest, different approaches have been developed to reduce the dMRI acquisition time while preserving the reconstruction quality.

We focus here on the problem of estimating white matter fiber orientations. A generalization to the study of additional microstructure parameters is beyond the scope of the present work. In this context, some of these methodologies are based on Spherical Deconvolution (SD) approaches, modeling the HARDI signal as a spherical convolution between the Fiber Orientation Distribution (FOD), representing the few active fiber orientations, and a kernel, representing the response signal of a single fiber \cite{Tournier2004, Alexander2005, Acqua2007}. Recently, SD has been utilized to recover the FOD from a reduced number of diffusion gradients (q-space under-sampling) \cite{Jian2007, Ramirez-Manzanares2007a,Tristan-Vega2011}. 

The FOD recovery problem can be formulated as an inverse problem, which is ill-posed and highly sensitive to noise. 
In this context, an efficient approach consists in leveraging the convex optimization theory by defining the FODs as a solution to a convex regularized minimization problem, incorporating \emph{a priori} information on the target FODs.
In particular, the coefficients of interest are imposed to be non-negative and real valued, since FODs are functions on the unit sphere expressing the orientation and the volume fraction of the fiber populations contained in a single voxel \cite{Tournier2007}. 

When expressed as set of discrete orientations, FODs are characterized by coefficients summing up to one in each voxel. 
Finally, leveraging the fact that, at the imaging resolution available currently, each single voxel is assumed to be populated by only a few fiber bundles, sparsity priors can be incorporated in the model (see, e.g., \cite{Ramirez-Manzanares2007a,Jian2007,Daducci2014b, Landman2012,Tristan-Vega2011,Auria2015,Michailovich2011a,Mani2015}).
In this context and inspired by the Compressed Sensing (CS) theory \cite{Donoho2006,Candes2006a}, many methods have obtained promising results recovering the FOD coefficients from a reduced number of diffusion gradients by leveraging sparsity-based priors.

{Constraint Spherical Deconvolution (CSD) \cite{Tournier2007} represents the first attempt to regularize the FOD recovery problem. Based on the fact that the FOD coefficients correspond to the fiber volume fractions, they are assumed to be non-negative. CSD proposes to iteratively solve a sequence of weighted $\ell_2$ norm problems whose weights, which depend on the FOD coefficients estimated at the previous iteration, are used to drive to zero negative and small coefficients. Inspired by the CS theory, ref. \cite{Landman2012} 
and \cite{Jian2007} suggested the use of the $\ell_1$ norm for the recovery of the volume fractions coefficients. However, ref. \cite{Daducci2014b} showed that the use of the $\ell_1$ norm, meant as simple sum of the coefficients value, is inconsistent with the fact that the volume fractions coefficients sum up to 1 by definition.

Additionally, ref. \cite{Daducci2014b} reformulated the FOD recovery in order to approach the $\ell_0$ minimization.
This is done by solving a sequence of weighted $\ell_1$ norm problems whose weights at each iteration correspond to the inverse of the values of the solution of the previous problem \cite{Candes2008a}. At convergence, these weights lead the values of the coefficients to be independent from the magnitude of the non-zero values, thus, mimicking the results of the $\ell_0$ pseudo-norm.

Ref. \cite{Ramirez-Manzanares2007a} proposed the use of spatial regularization to solve the SD ill-posed problem. In this work a piece-wise smoothness of the FOD coefficients is promoted while encouraging the coefficients to provide high contrast. Ref. \cite{Ramirez-Manzanares2007a} observed that the contribution of each FOD coefficient is highly correlated to the coefficients associated with its neighbor directions. They proposed to regularize the FOD recovery by penalizing the presence of coefficients that exhibit large variations along similar orientations. In addition, they provided a way to discard the noisy contributions by driving to zero the coefficients that are not sufficiently distant from the mean value of the coefficients of interest.}

In addition to the angular resolution, the spatial resolution of the acquired DW volumes is also important to achieve an accurate identification of the white matter paths. In principle, by considering small voxel sizes, the complexity of the inner fiber arrangements can be reduced. Thus, considering both high angular and spatial resolutions would be ideal to ensure an accurate fiber recovery \cite{Calabrese2014, Zhan2013, Vos2016}.

Currently, single-shot (SS) DW echo-planar imaging (EPI) is the most popular technique used to perform the dMRI acquisition in clinics. SS-DW-EPI enables extremely fast scan times, which make DW signals nearly immune to patient motion and mostly sensitive to the water molecule movements. The SS-DW-EPI scheme is typically applied in 2D, where each slice is sequentially excited, diffusion-encoded and then collected using a unique trajectory.
{Despite being very fast, 2D SS-DW-EPI is very inefficient due to the procedure required for the generation of intense diffusion gradients responsible for sensitizing the MRI signal produced by the examined tissues with respect to the diffusion phenomenon occurring along the gradient direction. Indeed, more than 50\% of the acquisition time is dedicated to the preparation of the diffusion gradients to be applied}, which, in the 2D setting, is required for each single slice.

A number of methods have been proposed in order to acquire DW signals more efficiently and devote the most of the acquisition time to achieve higher resolution and enhance the image quality \cite{Scherrer2016, Gao2014, Haldar2013}. Parallel imaging and 3D diffusion MRI acquisitions are some of them \cite{Jeong2003}. In particular, moving from the 2D to the 3D approach enables  reducing the number of times the diffusion preparation process is applied to cover the same brain volume. 

Ideally, a single signal preparation would be enough to encode an entire 3D DW volume. However, this is not the case in practice due to the fast decay DW signals are subject to. Consequently, 3D DW EPI acquisitions typically perform the collection of multiple sub-volumes, which still requires the signal preparation to be applied fewer times than in the 2D setting. In this context, 3D acquisitions can be used in dMRI to reduce the acquisition time required to scan each DW volume or, alternatively, achieve higher signals quality within the same scan time.

In order to accelerate the dMRI acquisition process, previously proposed q-space methods have mainly focused on reducing the number of diffusion gradients from fully sampled DW images. However, the recovery of white matter structures can further benefit from the incomplete acquisition of each k-space volume. 

Traditional approaches aiming at exploring the white matter microstructure from under-sampled kq-space data relies on two separate steps: first, DW images are recovered from incomplete k-space data, followed by the recovery of the fiber architecture from a limited number of q-space points.

Recently, methods recovering the diffusion profile directly from under-sampled kq-space data have started to gain significant attention \cite{Cheng2015,Mani2015,Magnotta2021,Ramos2020}.
{The work of \cite{Cheng2015} developed the first approach to simultaneously recover the diffusion
signal and the diffusion propagator from data sub-sampled in both the k and q 3D spaces. The regularization of the problem is performed both in the angular and in the spatial domain. More specifically, the proposed recovery problem promotes the sparsity of the DW images in the wavelet domain, the sparsity of the diffusion signal in the angular domain and the smoothness of the DW images. 

In the work of \cite{Cheng2015}, the dictionary used to sparsify the signal in the angular domain was generated by a Dictionary Learning technique that enabled the creation of an adaptive dictionary to fully characterize the signal in each single voxel.
The surfacelet transform has been shown to efficiently represent the directional information of the diffusion propagator by using only few coefficients. Inspired by these observations, ref. \cite{Sun2015} proposed to recover the full diffusion propagator from measurements that are under-sampled in kq-space by leveraging two sparsity priors. The first prior promotes sparsity of the diffusion propagator in the
surfacelet domain through the $\ell_1$ norm. The second prior promotes sparsity of the diffusion propagator in the gradient domain by means of the TV penalty.

In the work of \cite{Awate2013}, they used rotation-invariant dictionaries in which only few atoms are active by adapting their orientation to the diffusion MRI data. The recovery framework proposed by \cite{Awate2013} consists of alternating between the estimation of the coefficients identifying the dictionary atoms, the rotational transformation matrix, the phase contamination and the DW images. The sparsity of the vectors containing the coefficients individuating the dictionary atoms is leveraged to regularize the reconstruction. In addition, the corresponding DW images are recovered by promoting their sparsity in the wavelet domain.}

As the method presented in this paper, ref. \cite{Mani2015} focuses on the recovery of the FOD coefficients. In this work, the sparsity of the FOD coefficients is  promoted through the $\ell_1$ norm. Furthermore, a total variation (TV) penalty, acting on the DW images, indirectly promotes spatial fiber regularity within neighbor voxels.

In the present work, we develop a method to recover intra-voxel fiber configurations (FODs) at high spatio-angular resolution relying on a 3D kq-space under-sampling scheme. 
For each of a reduced number of q-space points, the time available for k-space sampling is used to acquire a sub-sampled k-space of a 3D sub-volume, rather than the full k-space of a 2D slice, thus, providing a potentially significant acceleration over the state-of-the-art. 

Anatomical constraints are leveraged in order to recover the FOD coefficients. 
First, prior knowledge of the spatial distribution of the brain tissues, which can be inferred from the image acquired in the absence of diffusion, is explicitly incorporated in the FOD recovery problem.
Secondly, the FOD estimate is defined as a solution to a regularized minimization problem, using the \textit{structured sparsity} prior proposed in \cite{Auria2015}. This regularization promotes simultaneously voxel-wise sparsity and spatial smoothness of fiber orientation.

Following \cite{Daducci2014b,Auria2015}, the resulting non-convex minimization problem is solved via a re-weighting scheme \cite{Candes2008a} involving a sequence of convex minimization problems with weighted sparsity prior. Specific to our contribution, a stochastic Forward--Backward (FB) algorithm is used to solve each of these convex problems, with convergence guarantees toward the minimum of the corresponding convex objective \cite{Combettes2016}.
One of the advantages of the stochastic approach is that it offers the possibility to handle efficiently multi-coil acquisitions involving large volumes of data, minimizing both the memory requirement per iteration and the reconstruction time.

Results from both simulated and real data experiments showed that the proposed approach outperformed, in terms of its potential of accelerating the acquisition while preserving the FOD reconstruction quality, both the existing method in kq-space \cite{Mani2015} and the traditional reconstruction approaches considering the TV prior for the reconstruction of the DW signals before the recovery of the FOD coefficients. 
Importantly, our experiments reveal that the optimal reduction of samples was achieved, not by simply reducing the number of q-space points while fully sampling the k-space, but rather by a combined kq-space under-sampling.

The remainder of the paper is organized as follows. 
The proposed approach is described in Section~\ref{Sec:M&M}. In particular, the FOD reconstruction inverse problem is described in Section~\ref{sec:PD}, and the proposed minimization problem and algorithm are given in \mbox{Section~\ref{sec:PA}.} 
In Section~\ref{sec:SR}, we provide the description of the experimental setup, and, in Section~\ref{sec:results}, we present the results obtained from both synthetic and real data. 
Finally, we discuss the achieved reconstruction performances, and we conclude in Section~\ref{sec:DC}.

\section{Materials and Methods} \label{Sec:M&M}
\subsection{Background}

\subsubsection{k and q Spaces Overview}\label{ssec:SD}
{The dMRI signal is captured by a multitude of MRI images, each of which is sensitive to  water molecule diffusion occurring at a specific angle. As for the standard MRI, the dMRI signal is spatially encoded
by the application of three spatial encoding gradients spanning the 3D space. The k-space
corresponds with good approximation to the 2D or 3D Fourier transform of the signal. In addition to the application of the spatial encoding
gradients, dMRI signals must be encoded for the diffusion process. 

To this aim,
additional gradients are applied during the acquisition, that are the diffusion gradient. Similarly to the k-space,
encoded by the spatial gradients, the q-space is the 3D space defined by the diffusion gradients applied during the dMRI acquisition process. Each dMRI volume is associated with diffusion along a specific direction $\hat{\textbf{q}}$ and, at a specific intensity $|\textbf{q}|$, characterizing a point $\textbf{q}=|\textbf{q}|\hat{\textbf{q}}$. The commonly used b-value parameter \cite{Tuch2004a} associated with each DW volume is defined as $b=4\pi^2|\textbf{q}|^4t$, with $t$ as the diffusion time.}

\subsubsection{FOD Recovery via SD Framework}\label{ssec:SD}

In this section, we describe the SD framework for FOD recovery, 
from q-space under-sampled measurements. We refer to \cite{Auria2015} for further details of the global FOD recovery problem.

Let $X \in \mathbb{R_+}^{(n+2)\times N}$ be the unknown matrix containing the FOD coefficients of interest, where $N$ is the number of imaged voxels and {$n$ is the number of dictionary atoms}. Each column of $X$ contains the $n+2$ coefficients of the corresponding voxel.
The objective is to find an estimate of the FOD field associated with all the voxels of the brain, from the degraded measurements $Y \in \mathbb{R}^{M \times N}$, given by $Y=\Phi X+\eta$, where $\Phi \in \mathbb{R}^{M \times (n+2)}$ is the observation matrix and $\eta \in \mathbb{R}^{M \times N}$ models the acquisition noise. Note that, for $\text{SNR}>2$  the acquisition noise consists in a realization of an additive random i.i.d. Gaussian \mbox{noise \cite{H.Gudbajartsson1995}}, where the Signal-to-Noise Ratio on the image acquired in absence of diffusion gradients (i.e., $\textbf{s}_0$ image) is defined as $\text{SNR}=\frac{\langle \textbf{s}_0 \rangle}{\sigma}$, where $\langle \cdot \rangle$ denotes the mean of its argument and $\sigma$ is the standard deviation of the noise.

More precisely, $\Phi$ is a known dictionary whose columns (called dictionary atoms) correspond to the response signals on $M$ q-points to $n$ single fiber orientation \cite{Ramirez-Manzanares2007a}. In addition, two atoms representing isotropic diffusion (typically in gray matter or CSF) are considered in the dictionary. The corresponding atoms are invariant under rotation in q-space.
The procedure adopted to generate $\Phi$ is provided in Section~\ref{ssec:dictionary}, based on the method developed in \cite{Daducci2014b}. In particular, the first row of $\Phi$ is dedicated to atom values in the absence of diffusion gradient, equal to 1: $(\Phi_{1,d})_{1 \leqslant d \leqslant (n+2)}=1$.

The rows of the data matrix $Y\in \mathbb{R}^{M \times N}$ span the unfolded diffusion volumes, acquired with $M$ gradients and normalized by the intensities of the volumes acquired in the absence of diffusion, denoted by $\textbf{s}_0$. The $\textbf{s}_0$ volume is assumed to be obtained from a separate acquisition and is used here as prior information. The first row of $Y$ is devoted to the normalized $\textbf{s}_0$ volume, i.e., all its coefficients are equal to $1$
.
Thus, for each voxel $v \in \{1,\ldots,N\}$, $Y_{1,v}=\sum_{i=1}^{n+2} \Phi_{1,i}X_{i,v}=\sum_{i=1}^{n+2}X_{i,v}=1$. 
In summary, introducing unitary lines in both $Y$ and $\Phi$ intrinsically forces the sum of the FOD coefficients to be equal to $1$, injecting directly this prior information into the inverse problem.

\subsection{Inverse Problem Formulation}\label{sec:PD}

\subsubsection{Proposed Measurement Model}\label{sec:model}

In this section, we describe the proposed under-sampled kq-space measurement model.
We consider acquisitions obtained from $C$ coil receivers, with $M$ diffusion gradients, sampling $K$ k-space points. The measurement matrix $\hat{Y}\in \mathbb{C}^{M C\times K}$ is expressed as follows:
\begin{eqnarray}
\hat{Y}=\left[ \hat{Y}_{q,c}\right]_{1\leqslant q\leqslant M, 1\leqslant c\leqslant C},\quad \text{with} \quad \hat{Y}_{q,c}=\mathcal{A}_{q,c} (X)+\eta_{q,c},
\label{model}
\end{eqnarray}
where $\hat{Y}_{q,c} \in \mathbb{C}^{1\times K}$ is the under-sampled k-space of the diffusion volume acquired with gradient indexed by $q$ and by the receiver coil indexed by $c$, and $\eta_{q,c} \in \mathbb{C}^{1\times K}$ is a realization of an independent and identically distributed (i.i.d) Gaussian noise. Indeed, while the noise contamination  on the magnitude signals is characterized by a Rician distribution, the original noise in k-space is Gaussian \cite{Henkelman1985}.
The matrix $X \in \mathbb{R}^{(n+2)\times N}$ represents the unknown FOD field of interest, and the linear operator $\mathcal{A}_{q,c}$ is given by:
\begin{equation}\label{model2}
\mathcal{A}_{q,c} (X)=\Phi_{q} X S_0U^{(c)}H^{(q,c)}FM^{(q)}.
\end{equation}
The various terms defining $\mathcal{A}_{q,c}$ are described as follows. 
$\Phi_q\in \mathbb{R}^{1\times (n+2)}$ is the $q${th} row of the dictionary $\Phi$ that spans the response of a single fiber oriented along $n$ different directions, to which two isotropic compartments are added, to represent the gray matter and the CSF. In order to implicitly force the FOD coefficients of each voxel to sum up to one, the Fourier coefficients of the $\textbf{s}_0$ volume are intentionally introduced in the first row of $\hat{Y}$. Here again, the $\textbf{s}_0$ volume is assumed to be obtained from a separate acquisition and is used as prior information. 

Measurements cannot be normalized as the diffusion volumes themselves are not accessible but only an incomplete k-space counterpart.
Thus, a matrix $S_0\in \mathbb{R}^{N\times N}$, whose diagonal elements correspond to the $\textbf{s}_0$ volume coefficients, is explicitly introduced in the full measurement operator.
The acquisition of the diffusion signal from multiple channels is taken into account through the diagonal matrix $U^{(c)}\in \mathbb{C}^{N\times N}$, which contains the sensitivity map of the corresponding receiver coil $c$. 
Moreover, motion and magnetic field inhomogeneities generate phase distortions that are taken into account in the diagonal matrix $H^{(q,c)}\in \mathbb{C}^{N\times N}$. 

We refer to Sections~\ref{ssec:coil} and \ref{ssec:phase} for the description of the procedures used to estimate the coil sensitivity and the phase contamination, respectively.
Finally, $F \in \mathbb{C}^{N \times N}$ represents the 3D Fourier matrix and $M^{(q)}\in \mathbb{R}^{N \times K_q}$ is a realization of a binary mask that under-samples each slice of the acquired volume. It is important to notice that different realizations of $M^{(q)}$ may be considered for each applied diffusion gradient. In addition, $M^{(1)}$ is chosen to be the identity matrix of $\mathbb{R}^N$, to fully acquire the Fourier coefficients of the $\textbf{s}_0$ signal (stored in $Y_{1,c}$) for normalization, segmentation and calibration purposes.\par

\subsubsection{Tissue Segmentation Constraints}

\begin{figure}[H]

\includegraphics[width=0.7\columnwidth]{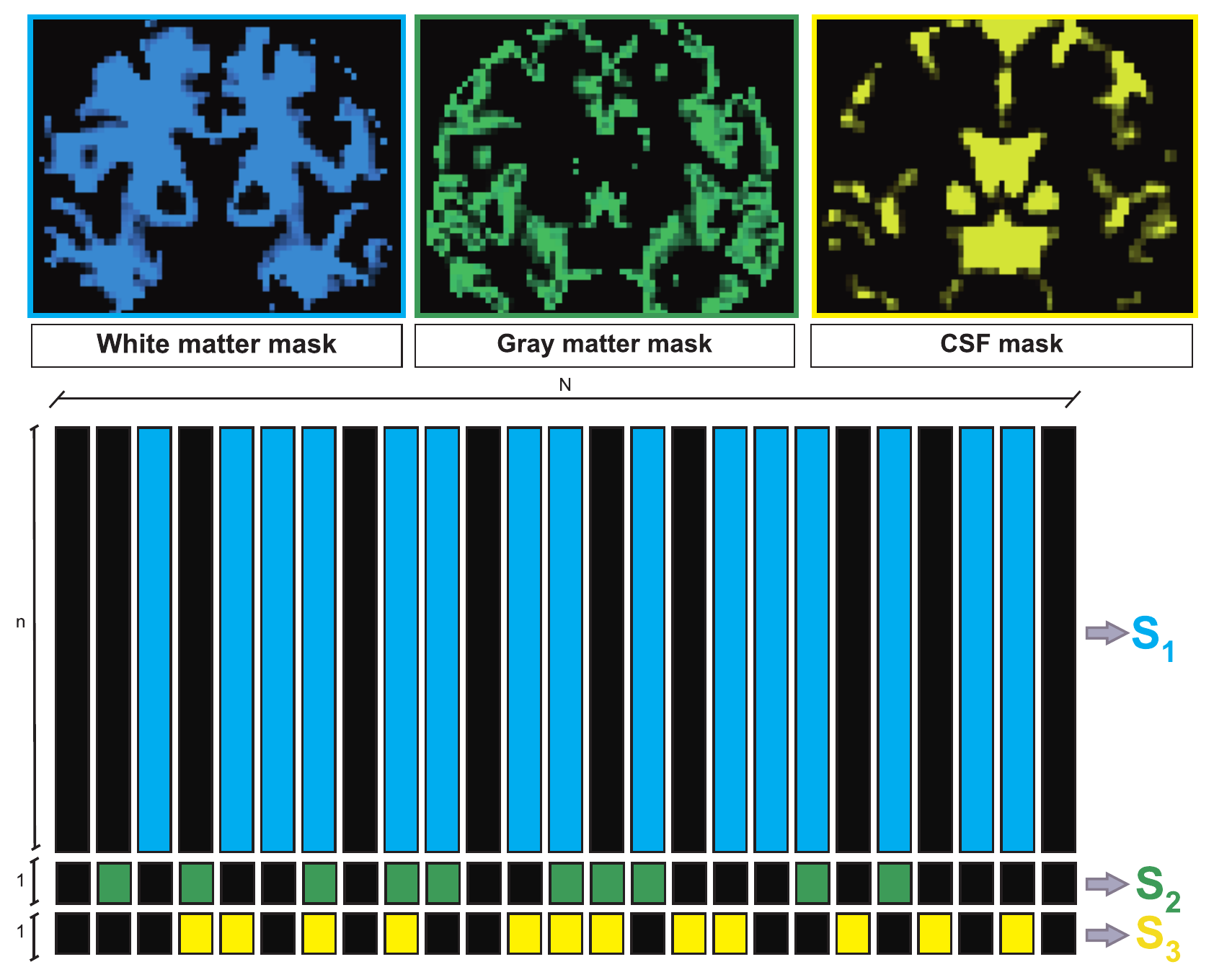}

\caption{Schematic representation of the matrices $S_1$, $S_2$ and $S_3$ involved in the FOD reconstruction. Each column of $S_1$ spans the FOD coefficients of the voxels containing white matter tissue (blue columns). Coefficients modeling gray matter and CSF are stored in $S_2$ (green columns) and in $S_3$ (yellow columns), respectively. The FOD reconstruction is performed by estimating $S_1$, $S_2$ and $S_3$, whose active columns are provided by the tissue segmentation maps (top).}
\label{fig:1}
\end{figure}

The spatial distribution maps of the different brain tissues can be obtained from the segmentation of a structural image of the imaged brain. Such images are generally acquired by default in clinical practice; alternatively, the spatial distribution maps can be estimated from the $\textbf{s}_0$ signal, which is always fully acquired for normalization and calibration purposes.
An illustration of a tissue distribution is given in the first row of Figure~\ref{fig:1}.
In our approach, we propose to explicitly take advantage of the prior knowledge of tissue distributions over the space in order to further regularize the ill-posed FOD estimation problem. 
We assume that the number of voxels containing white matter, gray matter and CSF tissues, are known and indicated by $N_1$, $N_2$ and $N_3$, respectively.

In order to fully exploit the tissue distribution information, we propose a novel formulation for the FOD global problem in the kq-space. 
To this aim, we define three different variables $S_1 \in \mathbb{R_+}^{n\times N_1}$, $S_2 \in \mathbb{R_+}^{1\times N_2}$ and $S_3 \in \mathbb{R_+}^{1\times N_3}$, related to the white matter tissue, the gray matter tissue and the CSF, respectively, as shown in Figure~\ref{fig:1}. 

More specifically, $S_1$ contains the effective FOD coefficients associated with the white matter fiber, $S_2$ models the isotropic behavior characterizing the gray matter tissue, and $S_3$ takes into account the isotropic behavior of the CSF. 
The object $X$ is fully characterized by $S=(S_1,S_2,S_3)$ through the linear mapping $X=\mathcal{Z}(S)$, where the operator $\mathcal{Z}$ concatenates the matrices resulting from the expansions of $S_1$, $S_2$ and $S_3$ with zero-valued columns in the places of the voxels that are known to not contain the corresponding tissue. 

A schematic representation of $X=\mathcal{Z}(S)$ is reported in Figure~\ref{fig:1}. In accordance with the white matter distribution maps (blue map on the top of Figure~\ref{fig:1}), the columns of $S_1$ (blue columns on the bottom) span the FOD coefficients of the voxels that are characterized by nonzero values in the map. In an analogous way, the coefficients of $S_2$ and $S_3$ (green and yellow columns on the bottom of Figure~\ref{fig:1}, respectively) represent the isotropic compartments associated with the voxels corresponding to the nonzero values in the gray matter and CSF maps (green and yellow map on the top of Figure~\ref{fig:1}, respectively).
Using the proposed notation, problem \eqref{model} can be rewritten as:
\begin{equation}\label{LIPkq}
\hat{Y}_{q,c}=\mathcal{A}_{q,c} \big(\mathcal{Z}(S)\big)+\eta_{q,c}.
\end{equation}
It is worth noticing that, in the particular case when the global FOD estimation formulation is applied only to white matter voxels, we have $X=S=S_1$, and $\mathcal{Z}$ is the identity operator.


\subsection{Minimization Problem and Algorithm}\label{sec:PA}

\subsubsection{Constrained Weighted-$\ell_1$ Minimization Problem}
\label{ssec:min_prob}

Recently, convex and non-convex optimization methods have gained a great deal of attention to solve ill-posed or ill-conditioned inverse problems. 
In this context, the estimate of the sought object is defined to be a {minimizer} of a sum of two functions: the data-fidelity term associated with the signal model, and the regularization term incorporating \emph{a priori} information of the target solution. 
On the one hand, when having an additive i.i.d. Gaussian noise, the data-fidelity of choice is the least squares criterion. On the other hand, sparsity-aware models, especially leveraging a CS framework \cite{Chen2001}, are particularly suitable for solving ill-posed or ill-conditioned inverse problems.

The most suitable way to promote sparsity consists in using the $\ell_0$ pseudo-norm, which counts for the number of non-zero coefficients of the signal of interest \cite{Donoho1995}. However, the $\ell_0$ pseudo-norm, being neither smooth nor convex, can be difficult to handle efficiently and it is often replaced by its convex relaxation, namely the $\ell_1$ norm \cite{Chen2001}. 
In the context of FOD reconstruction, the signal $X$ is sparse in the q-space as a consequence of the low number of fiber populations (i.e., non-zero FOD coefficients) that are expected to be contained within a voxel. 

However, when SD problems are considered, it was demonstrated in \cite{Daducci2014b} that the use of the $\ell_1$ norm is inconsistent with the physical constraint that the volume fractions of each voxel sum up to one. To overcome this difficulty, \cite{Daducci2014b} proposed to use a reweighting $\ell_1$ approach \cite{Candes2008a} to approximate the $\ell_0$ pseudo-norm in the context of the voxel-wise FOD estimation. Such an approach was subsequently generalized in \cite{Auria2015} to solve the inverse problem described in Section~\ref{ssec:SD}. This consists of solving sequentially several constrained weighted $\ell_1$ problems. The weights are chosen to simultaneously promote  voxel-wise sparsity and spatial smoothness of fiber orientation.
Formally, the problem is written as follows:
\begin{equation}\label{PROBLEM0}
\min_{\substack{X }} \| \Phi X - Y \|_2^2 
\text{ s.t. } X \in \mathcal{B}_{1,W}^+(\kappa),
\end{equation}
{where $\mathcal{B}_{1,W}^+(\kappa)$ denotes the intersection of the real positive orthant $ \mathbb{R}_+^{(n+2)\times N}$ with the weighted $\ell_1$ ball of radius ${\kappa \geqslant 0}$ centered in $0$ with weighting matrix $W = (W_{d,v})_{d,v} \in [0,+\infty[^{(n+2)\times N}$. Precisely, this is defined as}
\begin{equation}
\mathcal{B}_{1,W}^+(\kappa) = 
\big\{ X \in \mathbb{R}_+^{(n+2)\times N} \, \big| \, \| X \|_{1,W} \leqslant \kappa \big\},
\end{equation}
$\| \cdot \|_{1,W}$ is the weighted $\ell_1$ norm given by
\begin{equation}\label{weight1}
\| X \|_{1,W} = 
\sum_{v=1}^{N} \sum_{d=1}^{n+2}{W_{d,v}} | X_{d,v} |.
\end{equation}

In this work, we generalize this weighting scheme to solve the kq-space inverse 
 problem~\eqref{LIPkq}. In particular, we aim to solve
\begin{equation}
\underset{S = (S_1, S_2, S_3) }{\min}
\Big\| \mathcal{A} \big(\mathcal{Z}( S) \big) - \hat{Y} \Big\|_2^2 \text{ s.t. }
\begin{cases}
S_1 \in \mathcal{B}_{1,W}^+(\kappa), \\
S_2 \in \mathbb {R}^{1\times N_{2}}_+, \\
S_3 \in \mathbb {R}^{1\times N_{3}}_+,
\end{cases}
\label{PROBLEM}
\end{equation}
where $\mathcal{A} \colon  \mathbb{R}^{(n+2) \times N} \to \mathbb{C}^{M C \times k}$ consists of the concatenation of the operators $\mathcal{A}_{q,c}$, for all $(q,c)$.

\subsubsection{Algorithm for Constrained Weighted $\ell_1$ Minimization}

As proposed by the reweighting framework, problem~\eqref{PROBLEM} is solved several times, considering different weighting matrices $W$, in order to mimic the $\ell_0$ pseudo-norm. Note that the reweighting framework is underpinned by recent theoretical results \cite{Ochs2015, Geiping2018, Ochs2019, Repetti2019b} showing that sequentially minimizing convex problems with weighted-$\ell_1$ priors corresponds to solving for a critical point of a non-convex problem with a log-sum prior, itself a close approximation to the target $\ell_0$ prior. 
Considering one weighting cycle, a simple algorithm to solve problem~\eqref{PROBLEM} is the FB algorithm \cite{Combettes2005}. 

At each iteration $j \in \mathbb{N}$, it updates the variables $S^{(j+1)} = \big(S^{(j+1)}_1, S^{(j+1)}_2, S^{(j+1)}_3 \big)$ by performing a gradient step on the data-fidelity term, and three projection steps to handle the three constraints described in \eqref{PROBLEM}. 
However, computing the gradient step can be computationally unaffordable when processing high dimensional data, which is  particularly the case with multi-coil kq-space raw data.

To overcome this issue, we rely on a more evolved algorithm, namely the stochastic FB algorithm \cite{Combettes2016}, where only an approximation of the gradient of interest is computed at each iteration. Specifically, we propose to select randomly at each iteration a subset of the available coils, and approximate the gradient using only the data from the selected coils. The proposed method is described in Algorithm~\ref{algorithm1}.

Steps~\ref{algoFB:selection}--\ref{algoFB:gradapprox4} describe the computation of the approximated gradient. Below is an explanation of these steps. 
Let $\hat{Y}_c$ be the data measured by coil $c$ and $\mathcal{A}_c$ be its associated measurement operator. 
At iteration $j \in \mathbb{N}$, the gradient of the data fidelity term $ \| \mathcal{A} \big(\mathcal{Z}( S^{(j)}) \big) - \hat{Y} \|_2^2  $, denoted by $\overline{G}^{(j+1)}$, can be written as the sum $\sum_{c=1}^C G_c^{(j+1)}$, where $G_c^{(j+1)}$ corresponds to the gradient of $ \| \mathcal{A}_c \big(\mathcal{Z}( S^{(j)}) \big) - \hat{Y}_c \|_2^2  $. Using the stochastic approach described in \cite{Combettes2016}, the gradient $\overline{G}^{(j)}$ can be approximated by only updating a subset $\mathcal{S}_c^{(j)} \subset \{1, \ldots,C\}$ of the gradients $G_c^{(j)}$. More specifically, we define the approximation $\widetilde{G}^{(j)} $ of $\overline{G}^{(j)}$ as follows: 
$ \widetilde{G}^{(j)} = \sum_{c \in \mathcal{S}_c^{(j)}} G_c^{(j)} + \sum_{c \not\in \mathcal{S}_c^{(j)}} G_c^{(j)} $, 
where $G_c^{(j)} = \mathcal{Z}^\dagger \Big( \mathcal{A}_c^\dagger \big( \mathcal{A}_c (\mathcal{Z}(S^{(j)})) - \hat{Y}_c \big)\Big)$ if $c \in \mathcal{S}_c^{(j)}$, and $G_c^{(j)} = G_c^{(j-1)}$; otherwise, note that $(\cdot)^\dagger$ denotes the adjoint operator of its argument. 
Step~\ref{algoFB:gradapprox4} corresponds to a reformulation of this approximation.

\pagebreak

\begin{algorithm}[!h]
\caption{Stochastic FB to solve \eqref{PROBLEM}}
\label{algorithm1}
\begin{algorithmic}[1]
\State
\textbf{Input:} 
Let $S^{(0)} = (S_1^{(0)}, S_2^{(0)}, S_3^{(0)})$, with 
$S_1^{(0)}\in\mathbb{R_+}^{n\times N_1}$, 
$S_2^{(0)}\in\mathbb{R_+}^{1\times N_2}$, 
and $S_3^{(0)}\in\mathbb{R_+}^{1\times N_3}$. 
\State
Let $W \in\mathbb{R}_+^{n\times N_1}$, 
and $\left( \kappa,\gamma \right) \in \left[0,\infty \right[^2$.
\State
Let $G^{(0)}=\big( \mathcal{Z}^\dagger \Big(\mathcal{A}_c^\dagger \Big( \mathcal{A}_c \big(\mathcal{Z}(S^{(j)})\big)-\hat{Y}_c \Big)\Big) \big)_{1\leq c \leq C}$ and $ \displaystyle \widetilde{G}^{(0)}=\sum_{c=1}^C G_c^{(0)}$ 

\vspace*{0.1cm}

\State
\textbf{Iterations:}
\State
\textbf{For} {$j=0,1,2,\ldots$}
\State \label{algoFB:selection} 
$\quad \text{select randomly} \quad \mathcal{S}_c^{(j)} \subset \{1,..,C\} $
\State \label{algoFB:gradapprox1} 
\quad \textbf{For} $c= 1, \ldots, C$
\State \label{algoFB:gradapprox2}
{$\quad \quad G_c^{(j+1)} =
\begin{cases}
 \mathcal{Z}^\dagger \Big(\mathcal{A}_c^\dagger \Big( \mathcal{A}_c \big(\mathcal{Z}(S^{(j)})\big)-\hat{Y}_c \Big) \Big),
&   \text{if } c \subset \mathcal{S}_c^{(j)}  \\
 G_c^{(j)}
 &   \text{otherwise}
\end{cases}
$}
\State  \label{algoFB:gradapprox3}
\quad   \textbf{end for}
\State \label{algoFB:gradapprox4} 
$\quad  \widetilde{G}^{(j+1)}=\widetilde{G}^{(j)}-\sum_{c \in \mathcal{S}_c^{(j)}} (G_c^{(j)}-G_c^{(j+1)})$
\State  \label{algoFB:grad1}
\quad $\displaystyle \widetilde{S}^{(j)}= S^{(j)} -\gamma \widetilde{G}^{(j+1)}$
\State	\label{algoFB:proj1}
\quad $\displaystyle S_1^{(j)}=\mathcal{P}_{\mathcal{B}^+_{1,W}(\kappa)} \Big(\widetilde{S}_1^{(j)} \Big)$
\State	\label{algoFB:proj2}
\quad $\displaystyle S_2^{(j)}=\mathcal{P}_{\mathbb{R}_+^{1\times N_2}} \Big( \widetilde{S}_2^{(j)} \Big)$
\State	\label{algoFB:proj3}
\quad $\displaystyle S_{3}^{(j)}=\mathcal{P}_{\mathbb{R}_+^{1\times N_3}} \Big(\widetilde{S}_3^{(j)} \Big)$
\State
\textbf{end for}

\vspace*{0.1cm}

\State
\textbf{Output:} $S^\star = (S^\star_1, S^\star_2, S^\star_3)=\text{lim}_{j \to +\infty} (S_1^{(j)}, S_2^{(j)}, S_3^{(j)})$
\end{algorithmic}
\end{algorithm}

Steps~\ref{algoFB:proj1}--\ref{algoFB:proj3} describe the projection steps. The projection of a variable $x\in \mathbb{R}^N$ onto a nonempty, closed, convex set $C \subset \mathbb{R}^N$ is given by $\mathcal{P}_C(x)={\min_{y\in C}\| x-y \|_2^2}$. 
This corresponds to the closest point of $x$ belonging to $C$, using an Euclidean distance \cite{Combettes2005}. 
\mbox{Step~\ref{algoFB:proj1}} is the projection onto $\mathcal{B}_{1,W}^+(\kappa)$, related to the white matter.
In accordance with the $\ell_0$ approximation, $\kappa$ is chosen to represent the $S_1$ sparsity level.
Steps~\ref{algoFB:proj2} and \ref{algoFB:proj3} are the projections related to the gray matter and the CSF, respectively, performed onto $\mathbb{R}_+^{1 \times N_2}$ and $\mathbb{R}_+^{1 \times N_3}$. 

Let $(S^{(j)})_{j\in N}$ be a sequence generated by Algorithm~\ref{algorithm1}. 
According to~\cite{Combettes2005}, if $0<\gamma< 2 / \| \mathcal{A} \|_S^2$, where $\| \cdot \|_S$ denotes the spectral norm, then $(S^{(j)})_{j \in N}$ converges to a solution $S^\star$ of problem (\ref{PROBLEM}).
In our simulations, we consider that Algorithm~\ref{algorithm1} has converged when $\| S^{(j+1)} - S^{(j)} \|_2 < \nu \| S^{(j)} \|_2$, where $\nu =10^{-3}$ is a tolerance, fixed by the user.

\subsubsection{Weights Computation and Reweighting Procedure}

\textls[-15]{We propose to use an $\ell_1$ reweighting procedure to approximate the $\ell_0$ \mbox{pseudo-norm \cite{Candes2008a}.}} To this aim, the matrix of weights $W$ in Step~\ref{algoFB:proj1} of Algorithm~\ref{algorithm1}, is chosen using a similar approach as described in \cite{Auria2015}. 
The reweighting method is described in Algorithm~\ref{algorithm2}. 

\begin{algorithm}[!h]
\caption{Reweighting procedure}
\label{algorithm2}
\begin{algorithmic}[1]
\State
\textbf{Input:} 
Let $S^{(0)} = (S_1^{(0)}, S_2^{(0)}, S_3^{(0)})$, with 
$S_1^{(0)}\in\mathbb{R_+}^{n\times N_1}$, 
$S_2^{(0)}\in\mathbb{R_+}^{1\times N_2}$, 
and $S_3^{(0)}\in\mathbb{R_+}^{1\times N_3}$. 
\State
Let 
$W^{(0)} = \mathbf{1}_{n \times N_1}$.


\vspace*{0.1cm}

\State
\textbf{Iterations:}
\State
\textbf{For} {$t=0,1,\ldots,T-1$}

\State	\label{algoW:algoFB}
\quad $\displaystyle S^{(t+1)}=$ \textbf{Algorithm~\ref{algorithm1}}$\Big( S^{(t)}, W^{(t)}\Big)$


\State	\label{algoW:weights1}
\quad 
\textbf{compute} $W^{(t+1)}$ as per Equations \eqref{weight2} and \eqref{weight3}

\State
\textbf{end for}

\vspace*{0.1cm}

\State
\textbf{Output:} $S^{(T)} = \big( S^{(T)}_1, S^{(T)}_2, S^{(T)}_3 \big) $
\end{algorithmic}
\end{algorithm}

At each cycle $t \in \{0, \ldots, T-1\}$ of the reweighting scheme, problem~\eqref{PROBLEM} is solved using Algorithm~\ref{algorithm1} for a particular choice of the weighting matrix (see Step~\ref{algoW:algoFB}).
The weighting matrix $W^{(t+1)}=(W^{(t+1)})_{1 \le d \le n+2, 1 \le v \le N_1}$ is updated in Step~\ref{algoW:weights1} of Algorithm~\ref{algorithm2}, as follows:
\begin{equation}\label{weight2} 
W^{(t+1)}_{d,v}=
\frac{1}{ \tau^{(t+1)}+ B_{d,v}^{(t+1)}}, 
\end{equation}
where 
\begin{equation}\label{weight3} 
B_{d,v}^{(t+1)}
=\frac{1}{\vert \mathcal{N}(v)\vert}\sum\limits_{d'v' \in \mathcal{N}(dv)} \vert S_{1_{d'v'}}^{(t+1)} \vert,
\end{equation}  
and \begin{equation}
\begin{cases}
\tau^{(0)} = \operatorname{Var} \Big( B^{(0)} \Big),		\\
\tau^{(t+1)}= \max \big\{ \frac{\tau^{(t)}}{10},	 \overline{\tau} \big\},
\end{cases}
\end{equation}
with $\overline{\tau}>0$. 
The parameter $\tau^{(t)} $ is a stability parameter that avoids the weights  going to infinity when $B_{d,v}^{(t)}=0$. 
Essentially, $B^{(t)}$ is defined as the set of voxels in the support of $S_1^{(t)}$ that gives a blurred version of $S_1^{(t)}$. 
The \textit{spatial neighborhood} $\mathcal{N}(v)$ is defined as the group of voxels that share either a face, an edge or a vertex with the voxel of interest $v \in \{1, \ldots, N_1\}$. The \textit{angular neighborhood} $\mathcal{N}(d)$ is defined as the set of atoms associated with the directions that fall within a cone of 15$^\circ$ with the direction of interest $d$.
The neighborhood of an element individuated by the indices $d$, and $v$ is then indicated by $\mathcal{N}(dv)$ and corresponds to all the indices in the support of $S_1^{(t)}$ that are simultaneously included in $\mathcal{N}(v)$ and $\mathcal{N}(d)$. 

Weights designed in \cite{Auria2015} have  two-fold effects. 
Each element $W_{d,v}^{(t)}$ affects the corresponding element $S_{1_{d,v}}^{(t+1)}$ in such a way that large weights progressively force to zero spurious peaks, while small weights favor the presence of the FOD coefficients. Consequently the weighting matrix, associated with the $\ell_1$ norm, promotes sparsity.
In addition, the smoothness of the variation of the FOD coefficients across spatial neighborhoods is also promoted in \eqref{weight3} by averaging over neighbor voxels. Due to the discretization of the dictionary, fiber contributions are usually spread over a small angular support. 

That is the reason why \eqref{weight3} contains a summation (rather than averaging) over neighbor directions. If a spatio-angular neighborhood is reasonably homogeneous, the weight in (\ref{weight2}) will be reasonably small and the corresponding FOD coefficient will be preserved at the next iteration. On the contrary, spurious peaks isolated in their spatio-angular neighborhood will be associated with large weights. 
These large weights will tend to prevent such isolated peaks at the next iteration to not violate the weighted $\ell_1$-norm constraint.

Finally, the reweighting process stops when the number of allowed iterations $T$ is reached or there is no substantial relative variation between successive estimates of $S_1$, i.e., ${\Vert S_1^{(t+1)}-S_1^{(t)}\Vert}_2<10^{-3} {\Vert S_1^{(t+1)}\Vert}_2$.
Empirical observations suggest to choose $T=10$ \cite{Auria2015}. 

\subsubsection{Data Post-Processing}

Once the solution is found, a post-processing procedure is performed along the columns of $S_1$ in order to extract the directions of the fibers within each voxel. The identification of the highest peaks is performed among all the directions contained within a cone of $30^\circ$ for each different direction. No more than eight local peaks are assumed per voxel, and peaks smaller than 20\% of the maxima are disregarded in order to suppress spurious contributions \cite{Auria2015, Daducci2014b}.

\section{Experimental Setting}\label{sec:SR}

In this section, we provide the description of the different experimental settings. 
In particular, the specifications of the sampling schemes considered in the q- and in the k-spaces are provided in Sections~\ref{ssec:qsampling} and \ref{ssec:ksampling}, respectively. The method proposed to generate the dictionary is described in Section~\ref{ssec:dictionary}. Sections~\ref{ssec:phase} and \ref{ssec:coil} provide the procedures used to estimate the phase and the coil sensitivities, respectively.
A description of the metric used to quantitatively estimate the FOD reconstruction is provided in Section~\ref{ssec:eval}. 

\subsection{q-Space Under-Sampling}\label{ssec:qsampling}

\begin{figure}[H]

\includegraphics[width=0.8\columnwidth]{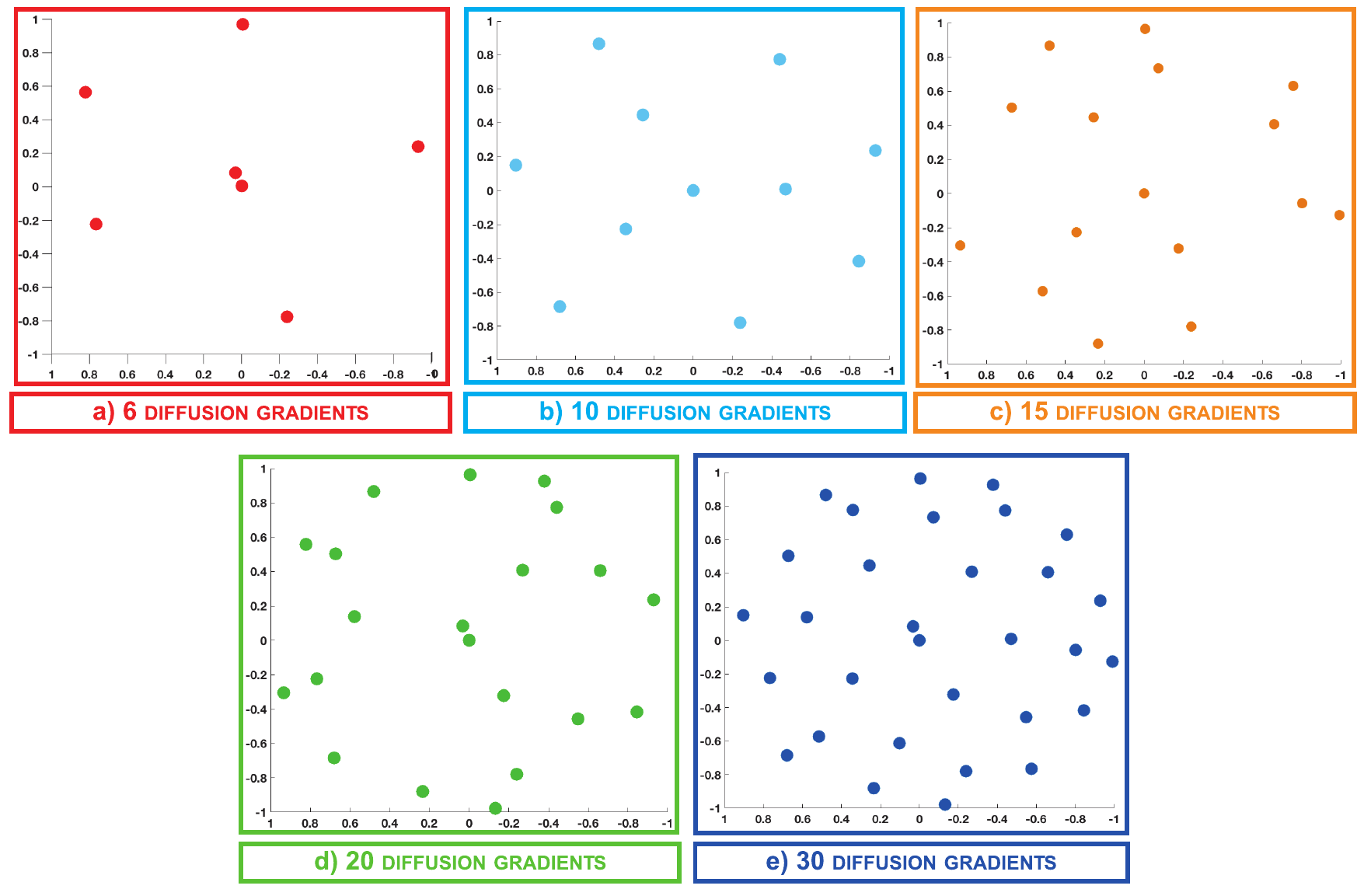}

\caption{Equatorial plane projection of the q-space under-sampling schemes considered for the synthetic data acquired with 6 (\textbf{a}), 10 (\textbf{b}), 15 (\textbf{c}), 20 (\textbf{d}) and 30 (\textbf{e}) diffusion gradients.}
\label{fig:4}
\end{figure}

In this section, we describe the scheme adopted for sampling the signal in q-space.
In the case of the synthetic data, the diffusion signal is uniformly sampled over a single shell using $30$ diffusion gradients with $b=1000\,\text{s/mm}^2$. In order to assess the FOD recovery in the presence of various kq-space regimes, the data provided was retrospectively under-sampled by selecting $6$, $10$, $15$ and $20$ uniformly distributed q-points, as reported in Figure~\ref{fig:4}.


In the case of the real data, the signal in q-space is sampled by $60$ gradients over \mbox{eight shells,} considering b-values going from $550$ to $4800\,\text{s/mm}^2$. In particular, $3$, $6$, $4$, $3$, $12$, $12$, $6$ and $13$ q-points are considered on sphere $1$ ($b=550\,\text{s/mm}^2$), sphere $2$ ($b=1050\,\text{s/mm}^2$), sphere $3$ ($b=1600\,\text{s/mm}^2$), sphere $4$ ($b=2150\,\text{s/mm}^2$), sphere $5$ ($b=2650\,\text{s/mm}^2$), sphere $6$ ($b=3200\,\text{s/mm}^2$), sphere $7$ ($b=4250\,\text{s/mm}^2$) and sphere $8$ ($b=4800\,\text{s/mm}^2$), respectively.

Three under-sampled datasets were created by retrospectively selecting the signal associated with $15$, $30$ and $45$ q-points in order to compare the FOD recovery in the presence of different q-space under-sampling regimes. Each dataset was created by retrospectively sampling q-points from different shells following a Gaussian distribution centered in the fifth shell, that is to say, q-points with b-values that ranges between $2000\,\text{s/mm}^2$ and $3000\,\text{s/mm}^2$ are sampled with higher probability \cite{Caruyer2013}. Five different realizations were created for each q-space setting.

\begin{figure}[H]

\includegraphics[width=0.85\columnwidth]{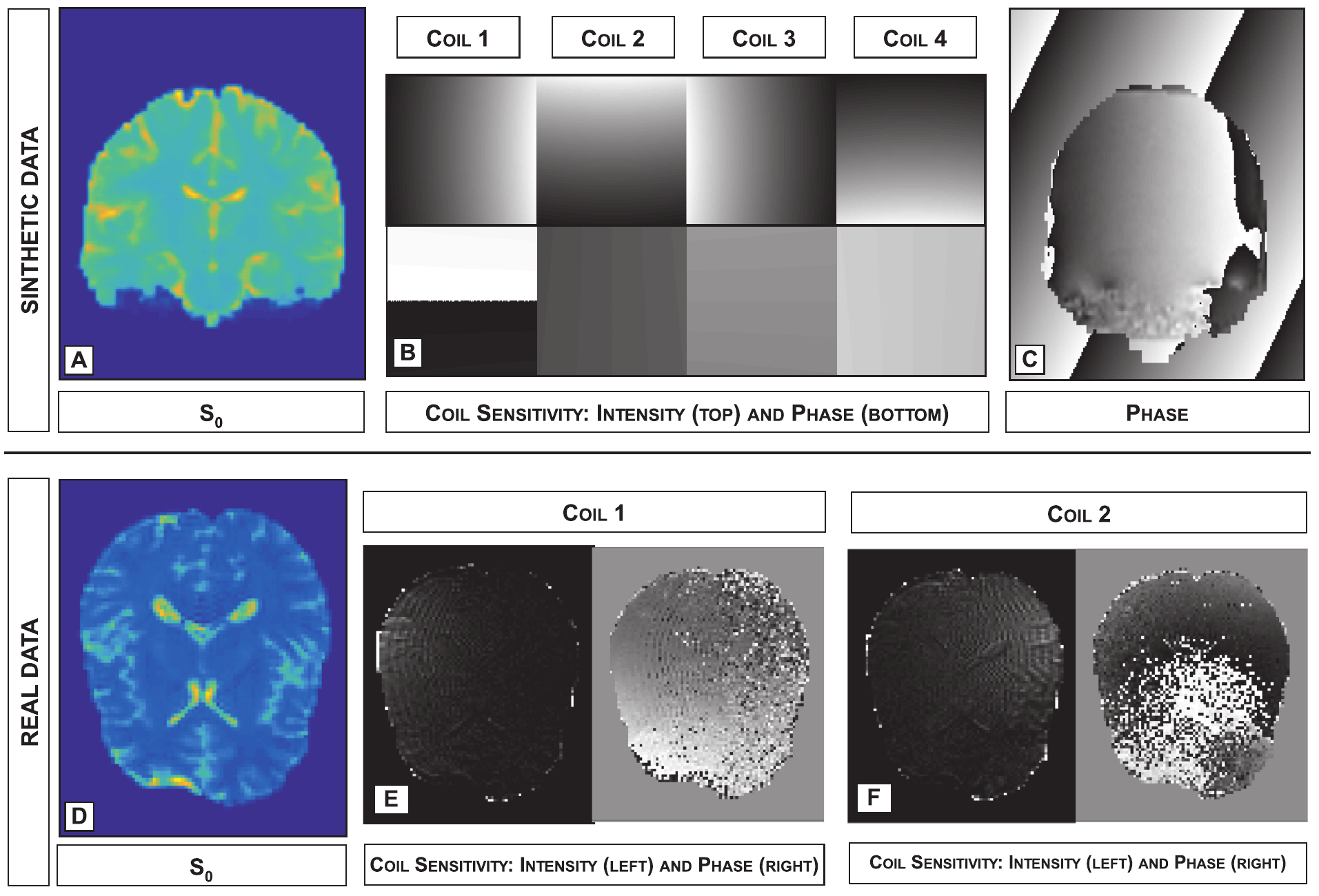}

\caption{
{First row: $s_0$ image (\textbf{A}) and coil sensitivity maps (magnitude (\textbf{top B}) and phase (\textbf{bottom B})) used in the experiments performed with synthetic data. (\textbf{C}) Example of phase map estimated with synthetic data when considering motion and magnetic field inhomogeneities.
Second row: $s_0$ image (\textbf{D}) and magnitude and phase estimated for two different coils (\textbf{E},\textbf{F}) for the experiments performed with real data.}}
\label{fig:8}
\end{figure}

For all the synthetic and real datasets, the acquisition of the signal in the absence of diffusion is required in order to obtain $\textbf{s}_0$.
Examples of this image, for both the synthetic and real data, are provided in Figure~\ref{fig:8}A,D, respectively.


\subsection{k-Space Under-Sampling}\label{ssec:ksampling}

As already explained, the objective of this work is to provide a robust reconstruction method when considering both q- and k-space under-sampling. The description of the schemes used to sample the signal in k-space is provided in this section.\par 
In both synthetic and real data, for each gradient $q \in \{1, \ldots, M\}$, selection masks $M^{(q)}$ are built such that the signal in k-space is sampled within a Cartesian grid by a continuous trajectory. We call \textit{k-space under-sampling factor} the ratio between the total amount of grid points in k-space, for a given field of view and resolution, and the number of samples considered for the reconstruction.

The potential of the proposed method is fully exploited in the 3D setting where the acquisition time can be significantly reduced by limiting both the number of excitations to cover each DW volume and the number of DW sub-volumes. As a proof of concept, the proposed FOD recovery method is evaluated for a specific class of 3D Fourier sampling patterns where the same 2D k-space under-sampling pattern is probed at all spatial frequencies in the third dimension of a selected sub-volume. By a simple inverse Fourier transform in this third dimension, the corresponding data can be cast, without loss of generality, as a slice-by-slice-identical 2D k-space under-sampling.
The scheme chosen for the 2D k-space sampling consists in fully sampling the central lines of the k-space while regularly skipping lines at the periphery, with the central zone of the k-space is required to estimate the phase affecting each shot (i.e., each sub-volume).

\begin{figure}[H]

\includegraphics[width=0.95\columnwidth]{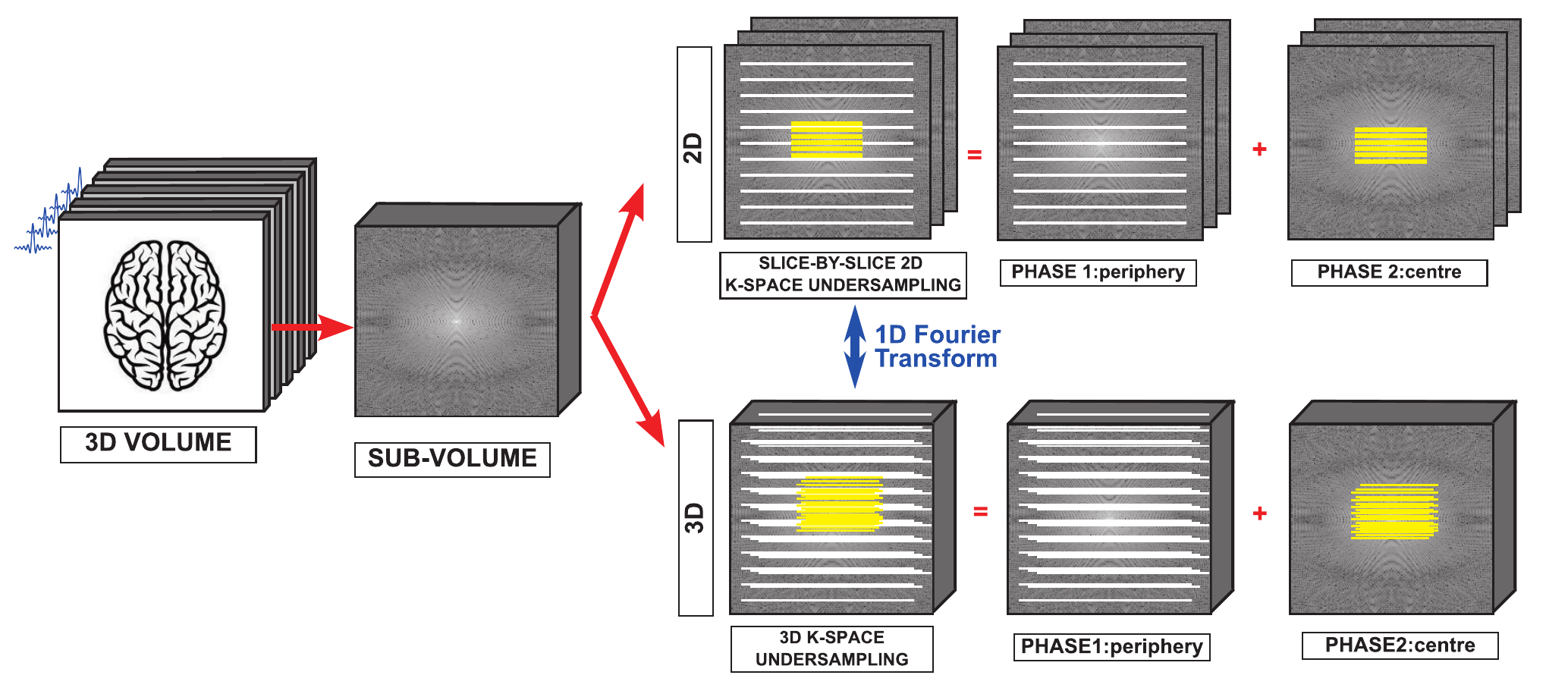}

\caption{
{Illustration of the acquisition of the 3D volume containing the brain. Each sub-volume is acquired considering a 3D two-phase EPI scheme (bottom line). The presented results were obtained considering a sequence of identical slice-by-slice 2D k-space under-sampling schemes (top line), which can be brought back to a specific class of 3D k-space under-sampling scheme to which a Fourier transform in the third dimension was applied. The proposed pattern results from the combination of two different regular schemes: the fully sampled central zone (yellow lines) and the regularly under-sampled periphery (white lines).}}
\label{fig:7}
\end{figure}

We argue that a two-phase EPI  acquisition protocol could be used to that effect. EPI commonly considers regularly sampled lines to enable a simple calibration of the eddy currents affecting each different k-space line. The proposed k-space under-sampling scheme could result from the combination of two uniform EPI schemes, one for the center and one for the periphery, with the eddy currents calibration critical only at the interface between the two regions. A schematic representation of the under-sampling scheme used with both synthetic and real data is reported in Figure~\ref{fig:7}. An experimental evaluation of a precise acquisition protocol is beyond the scope of the present work.


In the remainder of this work, all our experiments are designed with this k-space sampling scheme. Note, again without loss of generality, that the performance of the proposed approach is evaluated in a setting including a single DW (sub-)volume at each q-space point. Multiple sub-volumes would simply be acquired sequentially and reconstructed separately.

Note that the proposed method requires the complete k-space acquisition of the $\textbf{s}_0$ signal to implicitly force the FOD coefficients to sum up to one in each voxel. Furthermore, $\textbf{s}_0$ is used for normalization, segmentation and coil sensitivities calibration purposes as described in detail in Sections~\ref{ssec:phase} and \ref{ssec:coil}.

\subsection{Dictionary Generation}\label{ssec:dictionary}

In this section, we provide the description of the method used to create $\Phi$ for both the simulated and the real data. 
The elements of the dictionary $\Phi$ are generated by relying on the Gaussian Mixture Model of the q-space signal \cite{Tuch2002a}. 
More specifically, the dictionary $\Phi=\left[ \Phi_{q,d}\right]_{1\leqslant q\leqslant M, 1\leqslant d\leqslant (n+2)}$ is defined as $\Phi_{q,d}=\exp(-b \hat{\textbf{q}}^T \textbf{D}^{(d)} \hat{\textbf{q}})$, 
where $q$ is the index that explores the diffusion gradients $\textbf{q}$ (with b-value $b$ and orientation $\hat{\textbf{q}}$) considered for the acquisition, and $d$ is the index associated with the orientations chosen for the discretization of half of the unit sphere. We consider $n=500$ points for the discretization of the dictionary. 

Moreover, two additional atoms are considered to model the isotropic diffusion in the gray matter and CSF.
The $d$-th atom of the dictionary is indicated by $\left[\Phi_{q,d}\right]_{1\leq q\leq M}$, and this corresponds to the response of a single fiber, oriented along the $d$-th direction, subject to $M$ different diffusion gradients. 
The diffusion tensor $\textbf{D}=\textnormal{diag}(\lambda_1,\lambda_2,\lambda_3)$ characterizes each fiber population, where $\lambda_1$ is the longitudinal diffusivity, and $\lambda_2$ and $\lambda_3$ are the $2$ transverse diffusivity coefficients.
In particular, $\textbf{D}^{(d)}$ is the rotated version of $\textbf{D}$ along the direction $d$.

The diffusivity coefficients for the white matter fibers are fixed to the following values: $\lambda_1=17\times 10^{-4} \,\text{mm}^2/\text{s}$ and, $\lambda_2=\lambda_3=3\times 10^{-4} \,\text{mm}^2/\text{s}$ {\cite{Ramirez-Manzanares2007a,Jian2007}}.  
The diffusion tensors associated with the isotropic atoms are composed by 3 equal values in the diagonal. In particular, we consider $\lambda_1=\lambda_2=\lambda_3=17\times 10^{-4} \,\text{mm}^2/\text{s}$  and $\lambda_1=\lambda_2=\lambda_3=30\times 10^{-4} \,\text{mm}^2/\text{s}$ for the gray matter and CSF diffusivity coefficients, respectively.

\subsection{Phase Estimation}\label{ssec:phase}

In theory, diffusion images are real-valued. However, in practice, they are often contaminated by phase factors during the diffusion encoding process. This phase contamination is mostly due to magnetic field inhomogeneities and biological motion (e.g., physiological and involuntary patient motion).\par
Usually, methods dealing with signals directly in q-space overcome this difficulty by simply taking the modulus of the complex diffusion signal, in order to obtain real diffusion-weighted images. {However this method cannot be used in k-space and kq-space. Indeed, since the phase contamination breaks the Hermitian symmetry of the images, it cannot trivially be removed.} 
In particular, in the linear operator expressed in (\ref{model2}), the diagonal matrix $H^{(q,c)}$ takes into account this phase factor, for a fixed gradient $q$ and coil receiver $c$.\par

In the case of synthetic data, the phase contamination produced by the magnetic field inhomogeneities is provided within the contest data set.
In addition, effects due to motion are mimicked by adding a different linear phase map to the signal associated with each sub-volume of the phantom and coil receiver. The linear phase maps are generated in such a way that the corresponding k-space shift is constrained within 10 phase-encoding lines.
In each voxel $v \in \{1, \ldots,N\}$, the intensity of the signal in q-space is multiplied by $e^{i\theta_v}$, $\theta_v$ as the sum of the phases due to both motion and magnetic field inhomogeneities associated with $v$.
In Figure~\ref{fig:8}C, an example of estimated phase map due to motion and magnetic field inhomogeneities is provided.\par
For both synthetic and real data, phase-distortions are estimated from the central portion of the k-space data \cite{Miller2003,Pipe2002}, which is always fully sampled, (see Section~\ref{ssec:ksampling}). By performing the inverse Fourier transform of a zero-padded version of the fully sampled central part of the k-space, we obtain a complex-valued signal whose phase provides the diagonal elements of $H^{(q,c)}$.

\subsection{Coil Sensitivity Estimation}\label{ssec:coil}

In this section, we provide the method to estimate the coil sensitivity coefficients stored in the diagonal matrix $U^{(c)}$, appearing in the linear measurements described in \mbox{Equations \eqref{model} and \eqref{model2}} for both synthetic and real data. Images acquired in absence of diffusion from different coil receivers are combined through the \textit{sum of squares} \cite{Roemer1990} method in order to obtain a baseline image. Subsequently, the images associated with each receiver coil are divided by the baseline image in order to determine the corresponding sensitivity map.\par
For synthetic data, the acquisition from four different coil receivers was simulated by considering the toolbox at
 \url{http://bigwww.epfl.ch/algorithms/mri-reconstruction/}. Examples of the coil sensitivity, when $C=4$, are shown in Figure~\ref{fig:8}B.\par
Examples of the estimated sensitivity maps obtained in the real data experiment are provided in Figure~\ref{fig:8}E.

\subsection{Evaluation Criteria}\label{ssec:eval}

The quality of the fiber reconstruction was evaluated by using the metrics proposed in \cite{Auria2015}. For each single voxel, the fiber reconstruction evaluation takes into account the number of fibers correctly identified and the angular accuracy of the recovered direction. 

First, we define the \textit{success rate} (SR) index representing the proportion of voxels in which the number of fibers is correctly identified. More precisely, when all the estimated fibers fall within a tolerance cone of 30$^{\circ}$ around the true fibers, we have a success. The SR index depends on the \textit{false positive} and \textit{negative} rates that represent the average over all voxels of the number of overestimated and underestimated fibers per voxel.

Secondly, we define the \textit{mean angular error} $(\bar{\theta})$ as the average of the angular errors associated with each true fiber. The angular error is defined as $\theta=\frac{180}{\pi}arccos(\vert d_{true}\cdot d_{estimated}\vert)$,
where $d_{true}$ and $d_{estimated}$ are the true fiber direction and the direction of the closest estimated fiber, respectively.

\section{Results}\label{sec:results}

In this section, we present the results obtained when using the proposed method with both synthetic and real data. Specifically, in Sections~\ref{ssec:result_synthetic} and \ref{ssec:result_real} we discuss the results obtained for synthetic and real data, respectively.

\subsection{Synthetic Data}\label{ssec:result_synthetic}
A first analysis of the FOD reconstruction using the proposed kq-space under-sampling method was performed relying on the fiber configuration of the numerical phantom proposed in the \textit{ISMRM Tractography Challenge 2015} \cite{Hein2016}. 
The phantom consists of a volume of $N=90\times90\times108$ voxels, acquired by using $M=32$ diffusion gradients distributed over a single shell (see Section~\ref{ssec:qsampling} for more details).

We consider the acquisition from four coil receivers, where the sensitivity maps are simulated and estimated using the process provided in Section~\ref{ssec:coil}.
Phase contamination due to motion and magnetic field inhomogeneities was generated as described in Section~\ref{ssec:phase} and incorporated in the data.
The q-space signal was then converted to k-space through the Fourier transform and contaminated with Gaussian noise with zero mean and standard deviation $\sigma=\frac{\langle \textbf{s}_0\rangle}{\text{SNR}}$. Note that $SNR=30$ was chosen for the numerical simulations.
Lastly, selection masks, built as described in Section~\ref{ssec:ksampling}, were applied to the k-space of the diffusion-weighted images to obtain the under-sampled data in kq-space.
The dictionary $\Phi$ was generated using the procedure described in Section~\ref{ssec:dictionary}.

In the following section, the proposed recovery scheme is compared to three different reconstruction frameworks, which represent some of the state-of-the-art approaches for the FOD recovery in the presence of measurements under-sampled both in k and q-spaces separately and in kq-space jointly.
Among the state-of-the-art methods investigated, we can distinguish between two types of approaches: the two-step and the one-step approaches. 

The two-step approaches are the conventional approach to recover the FOD coefficients in the presence of under-sampled kq-space data, where the DW images are first reconstructed, and the FOD coefficients are subsequently estimated. 
As already mentioned in \mbox{Section~\ref{ssec:phase},} two-step approaches avoid to model the phase contamination due to motion and magnetic field inhomogeneities, by recovering complex images, whose imaginary part is discarded by taking the magnitude.
Among all the k-space methods, the TV-regularized approaches were considered for this comparison.
The one-step approaches consists on the fiber orientation estimation from the under-sampled kq-space signal directly. Contrary to the two-step approaches, the one-step approaches require the phase contamination to be modeled.
In particular, the proposed approach is a one-step method.
Below is a summary of the considered methods.

\begin{enumerate}

\item 
$TV-L2_+ $ is a two-step approach consisting in a first step where the DW volumes are recovered by using the $TV_+$ regularization and in a second step where FOD coefficients are recovered by relying on the non-negative least squares problem.
\begin{eqnarray}
&\textit{STEP 1:} \quad \underset{I\in \mathbb{C}^{M\times N}} {{\text{min}}}\| \mathcal{F}(I) - \hat{Y} \|_2^2+\lambda\|I\|_{TV},&
\end{eqnarray}
where $\hat{Y}\in \mathbb{C}^{MC\times N}$ contains the signal in kq-space.
Complex DW volumes are recovered in $I \in \mathbb{C}^{M\times N}$ from the TV regularized problem, where $\mathcal{F}(I) = (\mathcal{F}_{q,c}(I))$, with $\mathcal{F}_{q,c}(I)= I U^{(c)} F M^{(q)}$ for $q \in \{1,\ldots,M\}$ and $c \in \{1,\ldots,C\}$. To exclude the remaining phase contamination, the magnitude of the complex images is considered for the FOD recovery (i.e., $\tilde{I}=\frac{|I|}{S_0}$).  
\begin{eqnarray}
&\textit{STEP 2:}\quad \underset{X\in \mathbb{R}_+^{(n+2)\times N}} {{\text{min}}} \| \mathcal{D}(X) - \tilde{I} \|_2^2&
\end{eqnarray}
where $\mathcal{D}_{q}(X)=\Phi_q X$ for all $q \in \{1,\ldots,M\}$.
Positivity of the FOD coefficients is imposed in this step. 

\item $TV-STR_+$ is a two-step approach consisting in the recovery of both DW volumes and FOD coefficients from regularized problems. In the first step, the $TV$ prior is considered while, in the second step, the \textit{structured sparsity} prior proposed in \cite{Auria2015} is exploited.
\begin{eqnarray*}
&\textit{STEP 1:}\underset{I\in \mathbb{C}^{M\times N}} {{\text{min}}} \| \mathcal{F}(I) - \hat{Y} \|_2^2+\lambda\|I\|_{TV}&\\
&\textit{STEP 2:}\underset{X\in \mathbb{R}_+^{(n+2)\times N}}{{\text{min}}} \| \mathcal{D}(X) - \tilde{I} \|_2^2 \quad \text{s.t}\quad X \in \mathcal{B}_{1,W}^+(\kappa)&
\end{eqnarray*}

\item $TV_+$ is a one-step approach consisting in the recovery of the FOD coefficients from the kq-space signal applying the $TV$ prior to the images in order to implicitly promote a smooth variation of the FOD coefficients within neighbor voxels.
\begin{eqnarray*}
\underset{X\in \mathbb{R}_+^{(n+2)\times N}} {{\text{min}}} \| \mathcal{A}(X) - \hat{Y} \|_2^2+\lambda\|\Phi X \|_{TV}
\end{eqnarray*}
\item $STR_+-TISS$ is the proposed one-step approach consisting in the recovery of the FOD coefficients by solving the problem proposed in \eqref{PROBLEM} where the \textit{structured sparsity} prior and the spatial distribution of the different tissues is taken into account.

\end{enumerate}

The Primal-Dual \cite{Komodakis2014} and the FB algorithms were used to, respectively, solve the first and the second step of the two-step approaches. More specifically, in the second step of $TV-STR_+$, the FB algorithm was used multiple times to solve 10 weighted-$\ell_1$ problems. For all the frameworks, the optimal parameters were found using a grid-search approach. The evaluation of the image reconstruction in the case of the two-step approaches was performed relying on the SNR index. On the other hand, the SR index was chosen to evaluate the quality of the fiber orientation reconstruction and determine the best parameters for both the second step of the two-step approaches and the one-step approaches.

The parameters obtained in such a way are as follows: $1 e^{+3} \leq \lambda \leq 2.1 e^{+3}$ and $3.9 N \leq \kappa \leq 4.2N$. $TV_+$ is solved using the Primal-Dual algorithm, considering $4 e^{+5} \leq \lambda \leq 9.8 e^{+5}$. On the other hand, $STR_+-TISS$ was solved using the Algorithm~\ref{algorithm2}, with $\kappa= 4N_{wm}$, with $N_{wm}$ as the number of white matter voxels. However, in the case of the synthetic data, where only four coil receivers are considered, the gradient of the data fidelity term is fully computed and not approximated. Hence, at each iteration $j\in \mathbb{N}$, we choose $\mathcal{S}^{(j)}_c=\{1,2,3,4\}$ in Algorithm~\ref{algorithm1}, which reduces to the FB algorithm.

We  emphasize that additional tests were performed for $TV-L2_{+}$ and $TV-STR_{+}$ with $\lambda=0$, corresponding to remove the TV regularization for the DW images in the least-squares minimization problem. These tests have shown that the use of the TV prior provides better results. 

\begin{figure}[H]

\includegraphics[width=1\columnwidth]{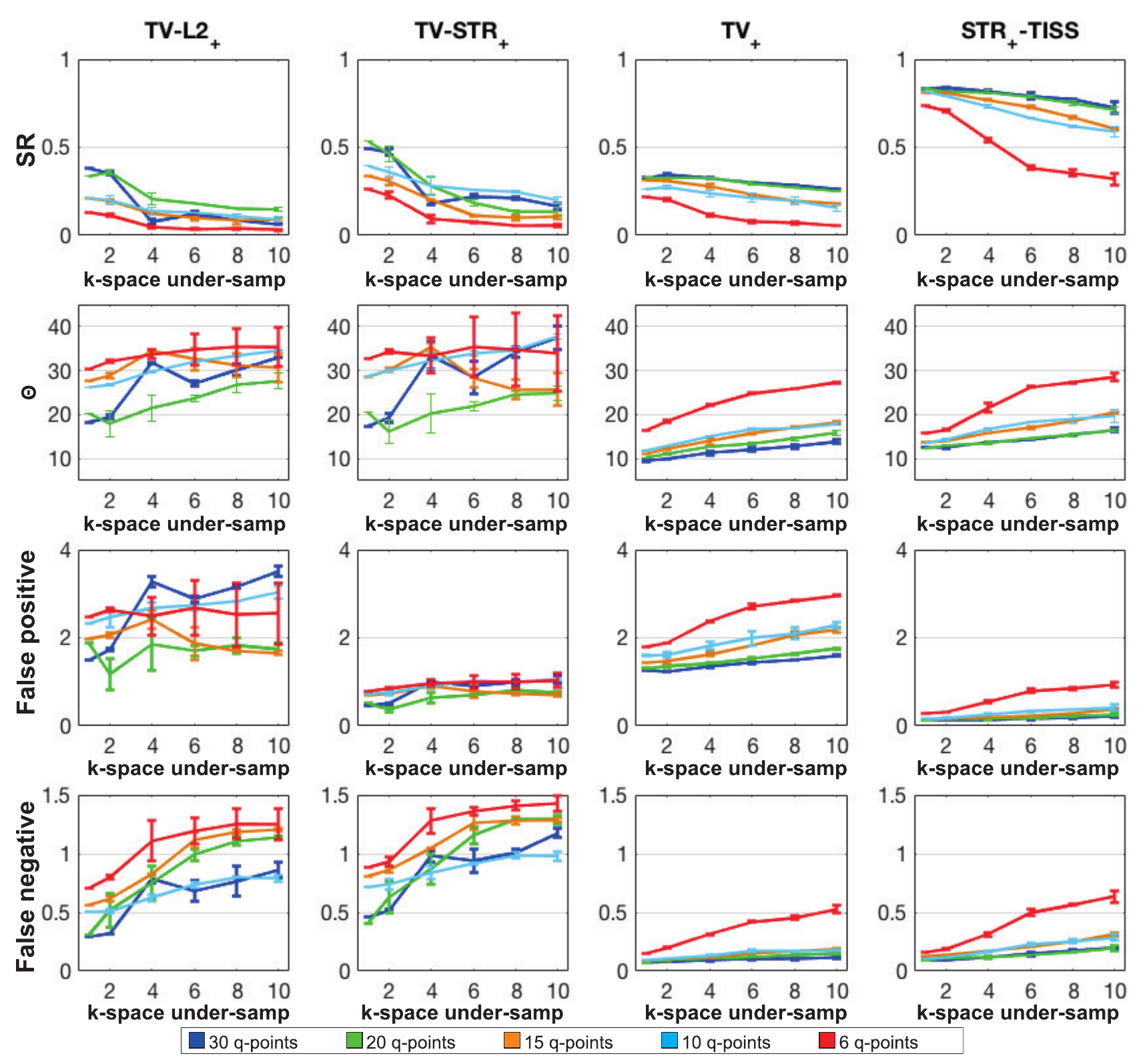}

\caption{Quantitative evaluation of the fiber orientation estimated by using four different methods for the synthetic data from the ISMRM Tractography Challenge 2015. For the evaluation, the SR (\textbf{first row}), the mean angular error (\textbf{second row}), the number of \textit{false positives} (\textbf{third row}) and \textit{false negatives} (\textbf{fourth row}) are shown for each different method along each column. Reconstructions were performed considering different number of q-points (i.e., $M \in \{6, 10, 15, 20, 30\}$), shown using different colors and for different \textit{k-space under-sampling factors} (i.e., $N/K \in \{1, \ldots, 10\}$).}
\label{fig:2}
\end{figure}

In Figure~\ref{fig:2}, the quantitative evaluation of the fiber reconstruction is provided in the case of various kq-space under-sampling regimes for the four different methods described above. Performances obtained considering different q-space under-sampling ratios are reported in different colors, while performances obtained considering different \textit{k-space under-sampling factors} are reported along the x-axis. 
The quality of the reconstruction was evaluated considering noisy data with $\text{SNR}=30$, and 5 different q-space under-sampling factors (i.e., $M\in \{6, 10, 15, 20, 30\}$), and $6$ different \textit{k-space under-sampling factors} ( i.e., $N/K \in \{$1$, $2$, $4$, $6$, $8$, $10$ \}$). 

The obtained SR index, mean angular error, and the rate of \textit{false positives} and \textit{negatives} are shown in Figure~\ref{fig:2}, from the first to the last row, respectively. The error bars correspond to the standard deviation of the evaluation indices obtained from k-space data corrupted by different noise realisations. 


Performances obtained with the two-step approaches are reported in the graphs in the first two columns of Figure~\ref{fig:2}. In the presence of k-space under-sampling regimes, both $TV-L2_+$ and $TV-STR_+$ are characterized by SR rates lower that the proposed $STR_+-TISS$ method. For $TV-L2_+$, the low performances result from the high number of spurious peaks that is estimated in each voxel as  can be observed in the diagram in the first column of Figure~\ref{fig:2}. When using the \textit{structured sparsity} prior to regularize the FOD coefficients in $TV-STR_+$ the number of \textit{false positive} drastically decreases, as shown in the corresponding graph in the second column of Figure~\ref{fig:2}. However, the number of \textit{false negative} moderately rises up, preventing the SR index to increase. 
In general, the direct regularization of the FOD provides better performances than $TV-L2_+$.

The last two columns of Figure~\ref{fig:2} give the graphs of the two one-step approaches: the $TV_+$ method (third column), and the proposed method $STR_+-TISS$ (fourth column). 
$TV_+$ does not reach high performances for any of the kq-space under-sampling regime. The SR index remains below $0.31$, that is reached considering $30$ q-points with no k-space under-sampling. These low SR values are due to the fact the $TV$ regularization is incapable of excluding the numerous spurious peaks that affect the FOD reconstruction. 

However, the performances obtained with the one-step approach $TV_+$, are shown to be quite robust to the k-space under-sampling. Indeed, the SR decreases only of $0.15$ (resp. $0.15$) and $\theta$ increases of $4^\circ$ (resp. $5^\circ$) going from no k-space under-sampling to a \textit{k-space under-sampling factor} of 10 with 30 q-points (resp. 10 q-points). 
As discussed in Section~\ref{ssec:min_prob}, and shown in \cite{Daducci2014b}, since fiber coefficients are implicitly forced to sum up to one in each voxel, imposing $\ell_1$ norm in this context is ineffective. Additional simulations  confirmed this behavior showing that using $\ell_1$ norm to regularize the FOD coefficients in addition to $TV$ did not lead to a significant difference in the results. 

Concerning the proposed $STR_+-TISS$ method,  $STR_+-TISS$ outperforms both the two-step and the one-step approaches here above presented. When compared to $TV_+$, $STR_+-TISS$ shows to substantially benefit from the direct regularization of the FOD coefficients, through the \textit{structured sparsity} prior.
Considering no k-space under-sampling with $6$-points (resp. $30$ q-points), SR $=0.31$ for $TV_+$ (resp. $0.48$), against SR $=0.84$ for $STR_+-TISS$ (resp. $0.86$).
For a \textit{k-space under-sampling factor} of 10 with $6$-points (resp. $30$ q-points), SR $=0.06$ for $TV_+$ (resp. $0.34$), against SR $=0.62$ for $STR_+-TISS$ (resp. $0.75$).
Thus $STR_+-TISS$ exhibits much higher performances when compared to $TV_+$. However, both methods are characterized by a similar amount of \textit{false negative}.

For the sake of brevity, we do not provide the comparison considering or ignoring the tissue spatial distribution. Additional simulations have shown that, when the regularization parameter $\kappa$ is known, FOD configurations recovered by the two methods are comparable. However, in practice $\kappa$ needs to be tuned, and using prior knowledge of the tissue distribution improves the reconstruction. This fact can be understood as follows. In the presence of under-estimated $\kappa$, coefficients associated with isotropic tissues compartments take precedence over the fiber compartments (since isotropic behavior can more easily fit the data). When the tissue distribution is taken into account, the coefficients associated with the fiber compartments are processed separately thus preventing them from being annihilated by the dominant presence of the isotropic compartment coefficients.

\begin{figure}[H]

\includegraphics[width=1\columnwidth]{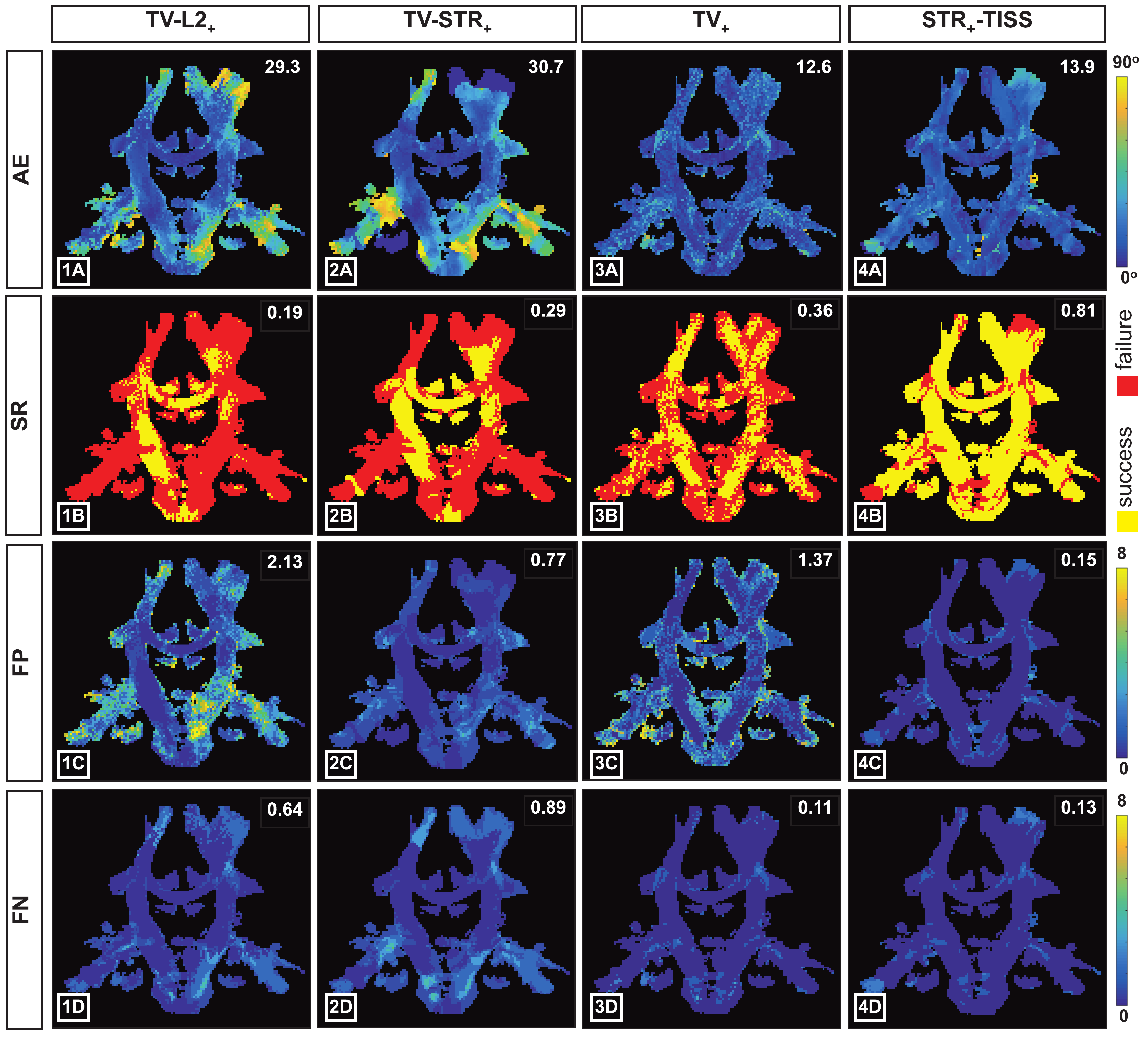}

\caption{Visualization of the quantitative results obtained by using the four different methods described in Section~\ref{ssec:result_synthetic}, applied to the synthetic data of the ISMRM Tractography Challenge 2015. Mean angular error (\textbf{1A}--\textbf{4A}), success rate (\textbf{1B}--\textbf{4B}), \textit{false positive} (\textbf{1C}--\textbf{4C}) and \textit{false negative} (\textbf{1D}--\textbf{4D}) maps obtained from the reconstructions considering M = $15$ q-points, with a \textit{k-space under-sampling factor} \mbox{N/K = $4$.} On the bottom right corner, values of the corresponding quantitative index are reported for the displayed slice.}
\label{fig:3}
\end{figure}

In Figure~\ref{fig:3}, we provide an illustrative example, considering $15$ diffusion gradients with a \textit{k-space under-sampling factor} of $4$, of the performance obtained from the four considered methods. In particular, we show the mean angular error (first row), the SR (second row), the \textit{false positive} (third row) and the \textit{false negative} (fourth row) maps. 
These results are confirming the quantitative results described in Figure~\ref{fig:2}. 
In particular, the maps evaluating the fiber configurations recovered from $TV-L2_+$ and $TV-STR_+$ show the worst performances. The SR maps in Figure~\ref{fig:3}(1B,2B) mostly show red pixels, indicating the presence of voxels with an uncorrected number of estimated fibers. 
For the $TV-L2_+$ method, the majority of such voxels is characterized by over-estimated fiber populations, as it is evidenced by the map in Figure~\ref{fig:3}(1C).

By comparing the maps in Figure~\ref{fig:3}(1C,2C), we can observe that the use of the \textit{structured sparsity} prior significantly decreases the amount of \textit{false positive}. Nevertheless, the map in Figure~\ref{fig:3}(2D) highlights the presence of higher \textit{false negatives} rates.
For the $TV_+$ method, the SR remains low, as it can be observed from the large areas with red pixels in Figure~\ref{fig:3}(3B). However, the angular accuracy of the estimated fiber configurations is significantly improved when compared to the two-step approaches (see maps on the first row of Figure~\ref{fig:3}).
In Figure~\ref{fig:3}(4B), the SR map obtained with the $STR-TISS_+$ method shows the higher rate of yellow pixels indicating the presence of a large number of voxels with the correct number of estimated fibers. In addition, the angular accuracy achieved with $STR-TISS_+$ remains low, as displayed by the map in Figure~\ref{fig:3}(4A).


\subsection{Real Data}\label{ssec:result_real}

The kq-space under-sampling framework was tested for the recovery of the fiber orientation from real data. The real data set consists of a volume of $100\times 100 \times 60$ voxels acquired on a {3T Magnetom Trio system (Siemens, Germany) with a pixel size of $2.2 \times 2.2 \times 2.2$ mm\textsuperscript{3}}. The dataset was acquired considering $17$ different coil receivers, and $60$ diffusion gradients distributed over multiple shells, as described in Section~\ref{ssec:qsampling}.
The complete signal was retrospectively under-sampled both in q- and k-space (as described in Sections~\ref{ssec:qsampling} and \ref{ssec:ksampling}, respectively) in order to compare the reconstruction performances considering different under-sampling regimes.

In the case of real data, $\Phi$ was generated by considering $n=500$ for the discretization of the sphere using the procedure described in Section \ref{ssec:dictionary}.
The tissue segmentation maps were estimated from the segmentation of the \textbf{$s_0$} signal and $H^{(q,c)}$ was estimated from the low resolution image as described in Section~\ref{ssec:phase}. The diagonal of the matrix $S_0$ is filled with the \textbf{$s_0$} values, which were recovered from the multi-coil signals while correcting for the estimated phase through a least squares approach. Ultimately, $U^{(c)}$ was estimated from the \textbf{$s_0$} signal as described in Section~\ref{ssec:coil}. Note that the real data was corrected for the EPI Nyquist ghosting prior to the reconstruction.

Results were obtained by solving problem~\eqref{PROBLEM} using Algorithm~\ref{algorithm2}, where the reweighing process is performed $10$ times at most, and $\kappa=5 \times N_{wm}$. The computation of the gradient in Algorithm~\ref{algorithm1} is approximated choosing 12 out of 17 coils at each iteration, i.e., $|\mathcal{S}_c|=12$. In particular, four of these coils are fixed (chosen at each iteration), and eight are randomly selected. In addition, a Nesterov acceleration \cite{Beck2009} is considered in Algorithm~\ref{algorithm1}. Although no theoretical result ensures the convergence of the resulting method, additional simulations have shown that Nesterov acceleration leads to much faster convergence while the same minimum is achieved using both the methods (with or without Nesterov acceleration).

\begin{figure}[H]

\includegraphics[width=1\columnwidth]{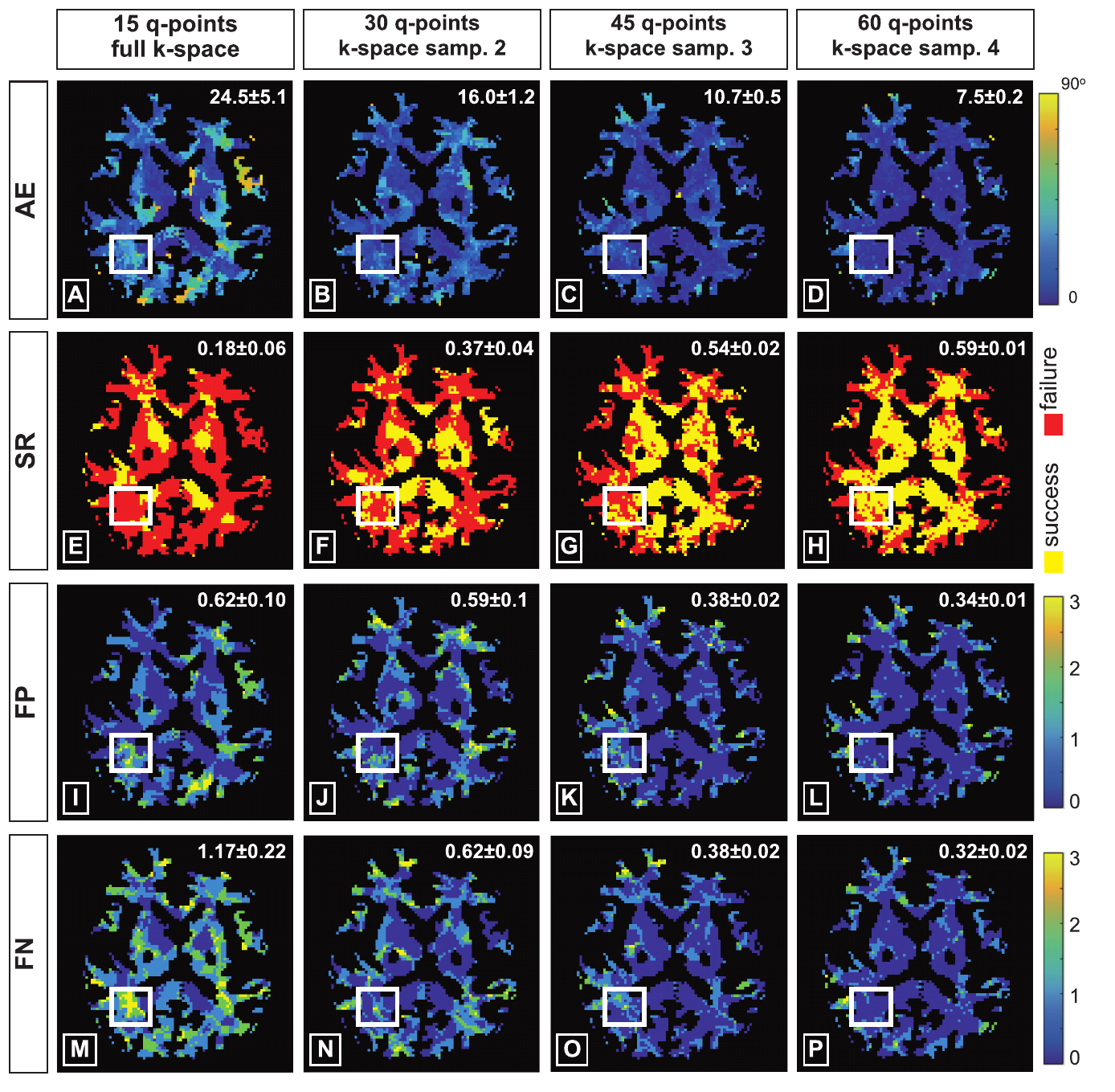}

\caption{Quantitative evaluation of the fiber orientation reconstruction performed on the real data by using the proposed kq-space under-sampling framework. Reconstructions considering 15 diffusion gradients with full k-space (\textbf{first column}), 30 diffusion gradients with \textit{k-space under-sampling factor} of 2 (\textbf{second column}), 45 diffusion gradients with \textit{k-space under-sampling factor} of 3 (\textbf{third column}) and 60 gradient with \textit{k-space under-sampling} factor of 4 (\textbf{fourth column})  were compared with the fiber configuration obtained using 60 q-points and full k-space. The maps of the mean angular error (\textbf{first row}), success rate (\textbf{second row}), \textit{false positives} (\textbf{third row}) and negatives (\textbf{fourth row}) are considered for the evaluation.}
\label{fig:6}
\end{figure}

Considering the fiber configuration recovered from 60 diffusion gradients with full k-space as reference, quantitative evaluation maps are computed and provided in Figure~\ref{fig:6}, for the fiber configurations obtained from four different kq-space under-sampling settings. For each different setting, the FOD reconstruction was performed 5 times, each considering a different realisation of q-space under-sampling. The results presented here corresponds to the mean and the standard deviation of the evaluation indices obtained considering different q-space under-sampling realizations.

In the first column of Figure~\ref{fig:6} the performances obtained considering $15$ diffusion gradients with full k-space is reported. The performances achieved with $30$ diffusion gradients considering a \textit{k-space under-sampling factor} of $2$ and $45$ diffusion gradients considering a \textit{k-space under-sampling factor} of $3$ are shown in the second and third columns of Figure~\ref{fig:6}, respectively. Ultimately, the evaluation maps of the fiber geometries recovered from $60$ diffusion gradients with \textit{k-space under-sampling factor} of $4$ are shown in the fourth column of Figure~\ref{fig:6}. 

The comparison between the evaluation maps provided in Figure~\ref{fig:6} shows that, considering the same overall kq-space under-sampling factor, the recovery from data that were under-sampled only in q-space results in lower quality reconstruction (SR $=0.18$ $\pm\,0.06$, $\bar{\theta}=24.5^\circ \pm 5.1$), as highlighted by the larger amount of red pixels and extended areas with high angular error in Figure~\ref{fig:6}A,E, respectively. On the other hand, the fiber configuration recovered from $60$ diffusion gradients with 4-fold k-space under-sampling exhibits the highest reconstruction quality (SR $=0.59$ $\pm\,0.01$, $\bar{\theta}=7.5^\circ \pm 0.2$) as  can be observed in the fourth column of Figure~\ref{fig:6}.  

\begin{figure}[H]
\widefigure
\includegraphics[width=0.9\textwidth]{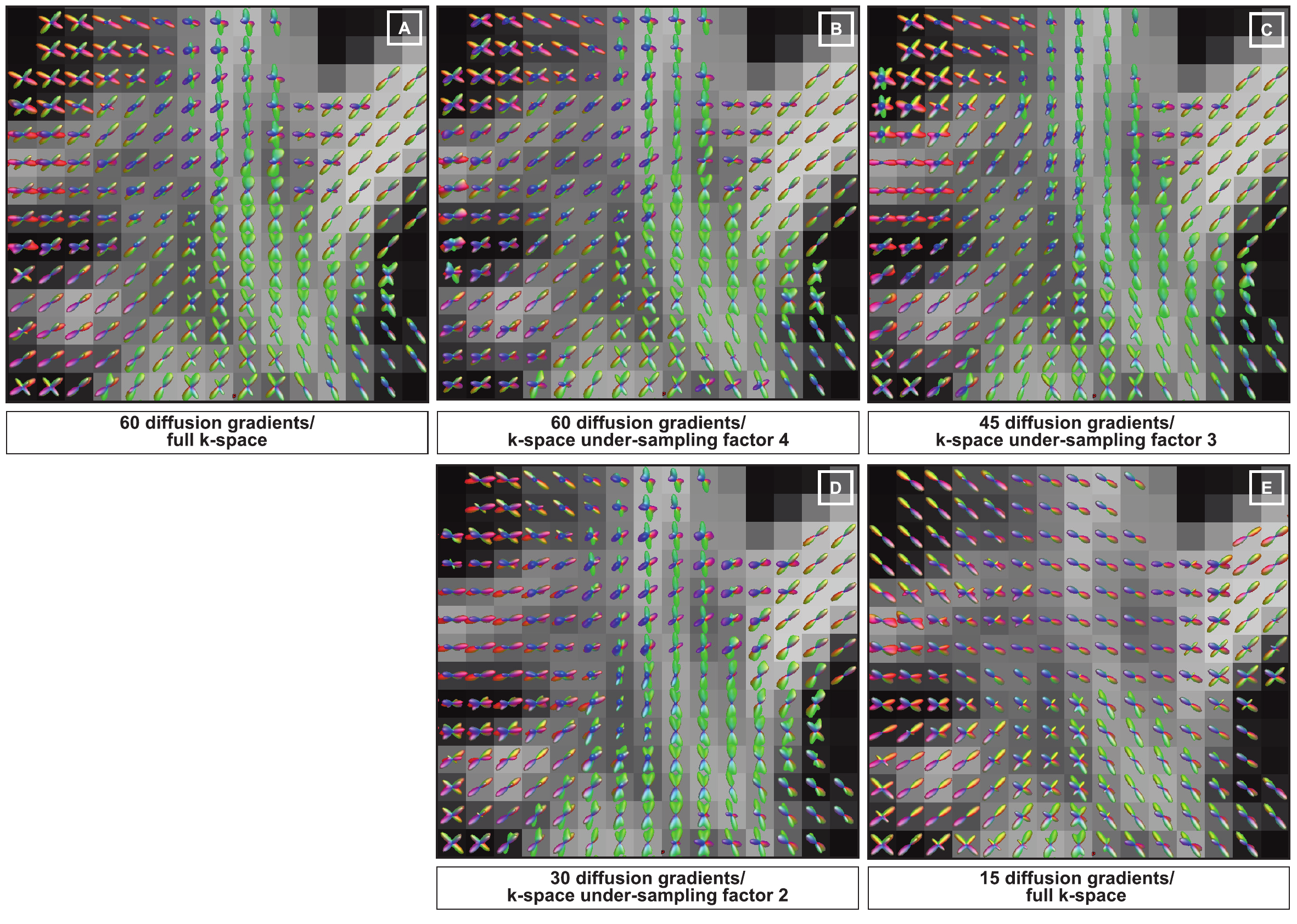}

\caption{Qualitative evaluation of the fiber orientation reconstruction performed on the real data by using the proposed kq-space under-sampling framework. The fiber configurations displayed correspond to the region framed highlighted by the white frame in Figure \ref{fig:6}. Reconstructions were performed considering 60 diffusion with full k-space (\textbf{A}), and with \textit{k-space under-sampling factor} of 4 (\textbf{B}), considering 45 diffusion gradients with \textit{k-space under-sampling factor} of 3 (\textbf{C}), 30 diffusion gradients with with \textit{k-space under-sampling factor} of 2 (\textbf{D}) and 15 diffusion gradients with full k-space (\textbf{E}).}
\label{fig:5}
\end{figure}

In Figure~\ref{fig:5}, we provide a zoomed view of the fiber configurations recovered from the real data in the presence of the same kq-space under-sampling settings described above.
The fiber configuration obtained considering $60$ diffusion gradients with full k-space is provided in Figure~\ref{fig:5}A as golden reference. Fiber geometries reported in \mbox{Figure~\ref{fig:5}B--D} appear to be very close to the golden reference. In contrast, the fiber configuration estimated in Figure~\ref{fig:5}E fails to represent the main fiber crossing revealed in the golden reference.

The analysis of the quantitative and qualitative results strongly suggests that strategies with combined kq-space sampling are advisable, when compared to q-space only strategies.


\clearpage
\end{paracol}
\nointerlineskip
\begin{paracol}{2}
\switchcolumn

In Table \ref{tab:1}, we compare the computational time required to process the real data with or without stochastic approach. Specifically, we investigate the computational time per iteration either when all the 17 coils are considered at each iteration (deterministic approach), or when 12 of the 17 coils are selected randomly at each iteration (stochastic approach). 
On the one hand, we can observe that the computational time per iteration when considering 12 coils per iteration corresponds (approximately) to $12/17$ times the computational time per iteration when considering 17 coils. This suggests that the computation of the gradient step in Algorithm~\ref{algorithm1} is much heavier than the computation of the projection step. 


\begin{specialtable}[H]
\caption{Comparison of the computational time required for the FOD recovery from the real data, considering four different under-sampling settings, using either 17 coils (deterministic approach) or 12 coils selected randomly per iteration (stochastic approach).}
\label{tab:1}

\setlength{\cellWidtha}{\columnwidth/6-2\tabcolsep+0.0in}
\setlength{\cellWidthb}{\columnwidth/6-2\tabcolsep+0.0in}
\setlength{\cellWidthc}{\columnwidth/6-2\tabcolsep-0.0in}
\setlength{\cellWidthd}{\columnwidth/6-2\tabcolsep-0.0in}
\setlength{\cellWidthe}{\columnwidth/6-2\tabcolsep-0.0in}
\setlength{\cellWidthf}{\columnwidth/6-2\tabcolsep-0.0in}
\scalebox{1}[1]{\begin{tabularx}{\columnwidth}{>{\PreserveBackslash\centering}m{\cellWidtha}>{\PreserveBackslash\centering}m{\cellWidthb}>{\PreserveBackslash\centering}m{\cellWidthc}>{\PreserveBackslash\centering}m{\cellWidthd}>{\PreserveBackslash\centering}m{\cellWidthe}>{\PreserveBackslash\centering}m{\cellWidthf}}
\toprule
\multicolumn{2}{c}{\textbf{Samp.}}&
\multicolumn{2}{c}{\textbf{17 Coils} \textbf{(Deterministic)}}
&\multicolumn{2}{c}{\textbf{12 Coils} \textbf{(Stochastic)}} \\
\midrule
\textbf{q} & \textbf{k} & \textbf{Time per Iter. (s)} & \textbf{Iter.} & \textbf{Time per Iter. (s)} & \textbf{Iter.} \\
\midrule
60 & 4 & 21.5 $\pm$ 0.13 & 3520 & 16.4 $\pm$ 0.15 & 3536 \\

45 & 3 & 17.9 $\pm$ 0.08 & 3603 & 13.5 $\pm$ 0.12 & 3618 \\

30 & 2 & 12.8 $\pm$ 0.07 & 3445 & 9.9 $\pm$ 0.13 & 3445 \\

15 & 1 & 7.7 $\pm$ 0.25 & 1856 & 6.1 $\pm$ 0.21 & 1854 \\
\bottomrule
\end{tabularx}}

\end{specialtable}

Thus, the computational time per iteration is lower using the stochastic approach than the deterministic approach. 
On the other hand, it can be noticed that both the deterministic and the stochastic approaches necessitate approximately the same number of iterations to reach the stopping criterion. 
Consequently, the use of a stochastic approach to recover the fiber configurations considering 12 coils per iteration enables to reduce both the memory requirement and the total computational time.

\section{Discussion}\label{sec:DC}

We developed a method to accelerate high angular and spatial resolution dMRI acquisition relying on a 3D kq-space under-sampling scheme. 
We provided a novel formulation to estimate the FOD coefficients from highly under-sampled sub-volumes. 
The proposed approach used two types of anatomical priors.
On the one hand, a brain tissue segmentation constraint, which does not required additional acquisitions and can be inferred from the no-diffusion-weighted image, was explicitly imposed for the FOD recovery. 
On the other hand,  \textit{structured sparsity},  developed previously in \cite{Auria2015}, was leveraged to promote simultaneously voxel-wise sparsity and spatial smoothness of fiber orientation.

The resulting minimization problem was approached via a reweighting scheme to solve a sequence of convex minimization problems, using a stochastic FB algorithm structure. By considering the stochastic variation, multi-coil data can be processed while minimizing both the memory requirement and the reconstruction time. Through synthetic and real data experiments, we demonstrated that the proposed recovery framework outperformed the existing method in kq-space and the traditional two-step reconstruction approaches recovering sequentially the diffusion weighted images and the FODs.  
Furthermore, we observed that, for equal overall under-sampling ratios, the proposed kq-space approach performed better when the k-space under-sampling was exploited rather than heavily under-sampling in q-space only.

Some limitations of our work are to be highlighted. First, a two-phase uniform EPI scheme was proposed to simultaneously fully sample the center and under-sample the periphery of the k-space data in order to minimize the effects arising from the irregular sampling. In this context, the use of more advanced models, taking into account the correction of the geometrical distortions typically affecting the EPI data and the use of acquisition schemes less prone to artifacts, would certainly contribute to improving the performance of the proposed method. 

Secondly, one of the critical steps of the proposed kq-space under-sampling method is the motion-induced phase estimation. We considered the phase estimated from the low resolution image associated with a fully sampled central k-space region. However, the use of more sophisticated methods to calibrate the motion-induced phase might further improve the performances of the kq-space approach and will be the focus of future investigations. 

Finally, future work should address the validation of our approach when additional microstructure parameters are considered, where the necessity for accelerating the acquisition is even stronger. The proposed approach generalizes naturally to this context, with the main difference lying in the definition of a larger and more general dictionary $\Phi$, accounting not only for fiber orientation, but also the diameter, etc.

\vspace{6pt} 



\authorcontributions{Conceptualization, Y.W., J.-P.T.; methodology, M.P., A.R., A.A., A.D., Y.W.; software, M.P.; validation, M.P.; formal analysis, M.P.; investigation, M.P.; resources, Y.W., J.-P.T.; data curation, M.P.; writing---original draft preparation, M.P., A.R., Y.W.;  writing---review and editing, M.P., A.R., A.D., Y.W.;  visualization, M.P.;  supervision, Y.W.; project administration, Y.W.; funding acquisition, Y.W. All authors have read and agreed to the published version of the manuscript.}

\funding{This work was supported by the UK Engineering and Physical Sciences Research Council (EPSRC, grants EP/M011089/1 and EP/T028270/1).}

\institutionalreview{The study was conducted according to the guidelines of the Declaration of Helsinki, and approved by the Ethics Committee of the Lausanne University Hospital (CHUV) and Centre for Biomedical Imaging (CIBM) as a pilot study (protocol code 20160429, date of approval 29 April 2016).}

\informedconsent{Informed consent was obtained from all subjects involved in the study.
}

\dataavailability{The data presented in this study are available on request from the corresponding author. The data are not publicly available due to size reasons. 
} 

\acknowledgments{The authors would like to thank the Center for Biomedical Imaging (CIBM) of the Geneva-Lausanne Universities for all the support received for the real data acquisition.}

\conflictsofinterest{The authors declare no conflict of interest.} 

\end{paracol}
\reftitle{References}


\begin{thebibliography}{999}

\bibitem[Jara(2013)]{jara2013theory}
Jara, Hernán
\newblock {\em {Theory of Quantitative Magnetic Resonance Imaging}}; World
  Scientific: Singapore, 
 2013.

\bibitem[Le~Bihan({2003})]{Bihan2003}
Le~Bihan, D.
\newblock {Looking into the functional architecture of the brain with diffusion
  MRI}.
\newblock {\em {Nat. Rev. Neurosci.}} {\bf {2003}}, {\em
  {4}},~{469--480}.

\bibitem[Sporns \em{et~al.}(2005)Sporns, Tononi, and K{\"{o}}tter]{Sporns2005}
Sporns, O.; Tononi, G.; K{\"{o}}tter, R.
\newblock {The human connectome: A structural description of the human brain}.
\newblock {\em PLoS Comput. Biol.} {\bf 2005}, {\em 1},~0245--0251.
\newblock
  doi:{\changeurlcolor{black}\href{https://doi.org/10.1371/journal.pcbi.0010042}{\detokenize{10.1371/journal.pcbi.0010042}}}.

\bibitem[Zhang \em{et~al.}(2009)Zhang, Schuff, Du, Rosen, Kramer,
  Gorno-Tempini, Miller, and Weiner]{Zhang2009}
Zhang, Y.; Schuff, N.; Du, A.T.; Rosen, H.J.; Kramer, J.H.; Gorno-Tempini,
  M.L.; Miller, B.L.; Weiner, M.W.
\newblock {White matter damage in frontotemporal dementia and Alzheimer's
  disease measured by diffusion MRI}.
\newblock {\em Brain} {\bf 2009}, {\em 132},~2579--2592.

\bibitem[Park \em{et~al.}(2004)Park, Westin, Kubicki, Maier, Niznikiewicz,
  Baer, Frumin, Kikinis, Jolesz, McCarley, and Shenton]{Park2004}
Park, H.J.; Westin, C.F.; Kubicki, M.; Maier, S.E.; Niznikiewicz, M.; Baer, A.;
  Frumin, M.; Kikinis, R.; Jolesz, F.A.; McCarley, R.W.; et al.
\newblock {White matter hemisphere asymmetries in healthy subjects and in
  schizophrenia: A diffusion tensor MRI study}.
\newblock {\em NeuroImage} {\bf 2004}, {\em 23},~213--223.

\bibitem[Callaghan \em{et~al.}(2000)Callaghan, Eccles, and Xia]{Callaghan2000}
Callaghan, P.T.; Eccles, C.D.; Xia, Y.
\newblock {NMR microscopy of dynamic displacements: k-space and q-space
  imaging}.
\newblock {\em J. Phys. E Sci. Instrum.} {\bf 2000}, {\em
  21},~820--822.
\newblock
  doi:{\changeurlcolor{black}\href{https://doi.org/10.1088/0022-3735/21/8/017}{\detokenize{10.1088/0022-3735/21/8/017}}}.

\bibitem[Basser \em{et~al.}(1994)Basser, Mattiello, and Le~Bihan]{Basser1994}
Basser, P.J.; Mattiello, J.; Le~Bihan, D.
\newblock {MR Diffusion Tensor Spectroscopy and Imaging}.
\newblock {\em Biophys. J.} {\bf 1994}, {\em 66},~259--267.

\bibitem[Tuch \em{et~al.}(2002)Tuch, Reese, Wiegell, Makris, Belliveau, and
  Wedeen]{Tuch2002a}
Tuch, D.S.; Reese, T.G.; Wiegell, M.R.; Makris, N.; Belliveau, J.W.; Wedeen,
  V.J.
\newblock {High angular resolution diffusion imaging reveals intravoxel white
  matter fiber heterogeneity}.
\newblock {\em Magn. Reson. Med.} {\bf 2002}, {\em 48},~577--582.

\bibitem[Wedeen \em{et~al.}(2005)Wedeen, Hagmann, Tseng, Reese, and
  Weisskoff]{Wedeen2005}
Wedeen, V.J.; Hagmann, P.; Tseng, W.Y.I.; Reese, T.G.; Weisskoff, R.M.
\newblock {Mapping complex tissue architecture with diffusion spectrum magnetic
  resonance imaging}.
\newblock {\em Magn. Reson. Med.} {\bf 2005}, {\em
  54},~1377--1386.

\bibitem[Tuch(2004)]{Tuch2004a}
Tuch, D.S.
\newblock {Q-ball imaging}.
\newblock {\em Magn. Reson. Med.} {\bf 2004}, {\em
  52},~1358--1372.

\bibitem[Zhang \em{et~al.}(2011)Zhang, Hubbard, Parker, and
  Alexander]{Zhang2011}
Zhang, H.; Hubbard, P.L.; Parker, G.J.M.; Alexander, D.C.
\newblock {Axon diameter mapping in the presence of orientation dispersion with
  diffusion MRI}.
\newblock {\em NeuroImage} {\bf 2011}, {\em 56},~1301--1315.

\bibitem[Zhang \em{et~al.}(2012)Zhang, Schneider, Wheeler-Kingshott, and
  Alexander]{Zhang2012}
Zhang, H.; Schneider, T.; Wheeler-Kingshott, C.A.; Alexander, D.C.
\newblock {NODDI: Practical in vivo neurite orientation dispersion and density
  imaging of the human brain}.
\newblock {\em NeuroImage} {\bf 2012}, {\em 61},~1000--1016.

\bibitem[Daducci \em{et~al.}(2014)Daducci, Canales-Rodriguez, Descoteaux,
  Garyfallidis, Gur, Lin, Mani, Merlet, Paquette, Ramirez-Manzanares, Reisert,
  Rodrigues, Sepehrband, Caruyer, Choupan, Deriche, Jacob, Menegaz, Prckovska,
  Rivera, Wiaux, and Thiran]{Daducci2014}
Daducci, A.; Canales-Rodriguez, E.J.; Descoteaux, M.; Garyfallidis, E.; Gur,
  Y.; Lin, Y.C.; Mani, M.; Merlet, S.; Paquette, M.; Ramirez-Manzanares, A.;
  et al.
\newblock {Quantitative comparison of reconstruction methods for intra-voxel
  fiber recovery from diffusion MRI}.
\newblock {\em IEEE Trans. Med Imaging} {\bf 2014}, {\em
  33},~384--399.

\bibitem[Ning \em{et~al.}(2015)Ning, Laun, Gur, DiBella, Deslauriers-Gauthier,
  Megherbi, Ghosh, Zucchelli, Menegaz, Fick, St-Jean, Paquette, Aranda,
  Descoteaux, Deriche, O'Donnell, and Rathi]{Ning2015}
Ning, L.; Laun, F.; Gur, Y.; DiBella, E.V.; Deslauriers-Gauthier, S.; Megherbi,
  T.; Ghosh, A.; Zucchelli, M.; Menegaz, G.; Fick, R.; \mbox{et al.}
\newblock Sparse Reconstruction Challenge for diffusion MRI: Validation on a
  physical phantom to determine which acquisition scheme and analysis method to
  use?
\newblock {\em Med. Image Anal.} {\bf 2015}, {\em 26},~316--331.

\bibitem[Tournier \em{et~al.}(2004)Tournier, Calamante, Gadian, and
  Connelly]{Tournier2004}
Tournier, J.D.; Calamante, F.; Gadian, D.G.; Connelly, A.
\newblock {Direct estimation of the fiber orientation density function from
  diffusion-weighted MRI data using spherical deconvolution}.
\newblock {\em NeuroImage} {\bf 2004}, {\em 23},~1176--1185.

\bibitem[Alexander(2005)]{Alexander2005}
Alexander, D.C.
\newblock {Maximum entropy spherical deconvolution for diffusion MRI.}
\newblock {\em Inf. Process. Med. Imaging} {\bf 2005}, {\em
  19},~76--87.

\bibitem[Acqua \em{et~al.}(2007)Acqua, Rizzo, Scifo, Clarke, Scotti, and
  Fazio]{Acqua2007}
Acqua, F.D.; Rizzo, G.; Scifo, P.; Clarke, R.A.; Scotti, G.; Fazio, F.
\newblock {A Model-Based Deconvolution Approach to Solve Fiber Crossing in
  Diffusion-Weighted MR Imaging}.
\newblock {\em IEEE Trans. Biomed. Eng.} {\bf 2007}, {\em
  54},~462--472.

\bibitem[Jian and Vemuri(2007)]{Jian2007}
Jian, B.; Vemuri, B.C.
\newblock {A unified computational framework for deconvolution to reconstruct
  multiple fibers from diffusion weighted MRI}.
\newblock {\em IEEE Trans. Med. Imaging} {\bf 2007}, {\em
  26},~1464--1471.
\newblock
  doi:{\changeurlcolor{black}\href{https://doi.org/10.1109/TMI.2007.907552}{\detokenize{10.1109/TMI.2007.907552}}}.

\bibitem[Ramirez-Manzanares \em{et~al.}(2007)Ramirez-Manzanares, Rivera,
  Vemuri, Carney, and Mareei]{Ramirez-Manzanares2007a}
Ramirez-Manzanares, A.; Rivera, M.; Vemuri, B.C.; Carney, P.; Mareei, T.
\newblock {Diffusion basis functions decomposition for estimating white matter
  intravoxel fiber geometry}.
\newblock {\em IEEE Trans. Med. Imaging} {\bf 2007}, {\em
  26},~1091--1102.

\bibitem[Trist{\'{a}}n-Vega and Westin(2011)]{Tristan-Vega2011}
Trist{\'{a}}n-Vega, A.; Westin, C.F.
\newblock {Probabilistic ODF estimation from reduced HARDI data with sparse
  regularization}.
\newblock In {\em International Conference on Medical Image Computing and Computer-Assisted Intervention}; Springer: Berlin/Heidelberg, Germany, {2011}; {Volume 6892 LNCS}, pp. 182--190.

\bibitem[Tournier \em{et~al.}(2007)Tournier, Calamante, and
  Connelly]{Tournier2007}
Tournier, J.D.; Calamante, F.; Connelly, A.
\newblock {Robust determination of the fibre orientation distribution in
  diffusion MRI: Non-negativity constrained super-resolved spherical
  deconvolution}.
\newblock {\em NeuroImage} {\bf 2007}, {\em 35},~1459--1472.

\bibitem[Daducci \em{et~al.}(2014)Daducci, {Van De Ville}, Thiran, and
  Wiaux]{Daducci2014b}
Daducci, A.; {Van De Ville}, D.; Thiran, J.P.; Wiaux, Y.
\newblock {Sparse regularization for fiber ODF reconstruction: From the
  suboptimality of $\ell_2$ and $\ell_1$ priors to $\ell_0$}.
\newblock {\em Med. Image Anal.} {\bf 2014}, {\em 18},~820--833.

\bibitem[Landman \em{et~al.}(2012)Landman, Bogovic, Wan, ElShahaby, Bazin, and
  Prince]{Landman2012}
Landman, B.A.; Bogovic, J.A.; Wan, H.; ElShahaby, F.E.Z.; Bazin, P.L.; Prince,
  J.L.
\newblock {Resolution of crossing fibers with constrained compressed sensing
  using diffusion tensor MRI}.
\newblock {\em NeuroImage} {\bf 2012}, {\em 59},~2175--2186.
\newblock
  doi:{\changeurlcolor{black}\href{https://doi.org/10.1016/j.neuroimage.2011.10.011}{\detokenize{10.1016/j.neuroimage.2011.10.011}}}.

\bibitem[Aur{\'{i}}a \em{et~al.}(2015)Aur{\'{i}}a, Daducci, Thiran, and
  Wiaux]{Auria2015}
Aur{\'{i}}a, A.; Daducci, A.; Thiran, J.P.; Wiaux, Y.
\newblock {Structured sparsity for spatially coherent fibre orientation
  estimation in diffusion MRI}.
\newblock {\em NeuroImage} {\bf 2015}, {\em 115},~245--255.

\bibitem[{Michailovich} \em{et~al.}(2011){Michailovich}, {Rathi}, and
  {Dolui}]{Michailovich2011a}
{Michailovich}, O.; {Rathi}, Y.; {Dolui}, S.
\newblock Spatially Regularized Compressed Sensing for High Angular Resolution
  Diffusion Imaging.
\newblock {\em IEEE Trans. Med. Imaging} {\bf 2011}, {\em
  30},~1100--1115.
\newblock
  doi:{\changeurlcolor{black}\href{https://doi.org/10.1109/TMI.2011.2142189}{\detokenize{10.1109/TMI.2011.2142189}}}.

\bibitem[Mani \em{et~al.}(2015)Mani, Jacob, Guidon, Magnotta, and
  Zhong]{Mani2015}
Mani, M.; Jacob, M.; Guidon, A.; Magnotta, V.; Zhong, J.
\newblock {Acceleration of high angular and spatial resolution diffusion
  imaging using compressed sensing with multichannel spiral data}.
\newblock {\em Magn. Reson. Med.} {\bf 2015}, {\em 73},~126--138.

\bibitem[Donoho(2006)]{Donoho2006}
Donoho, D.L.
\newblock {Compressed sensing}.
\newblock {\em IEEE Trans. Inf. Theory} {\bf 2006}, {\em
  52},~1289--1306.

\bibitem[Cand{\`{e}}s(2006)]{Candes2006a}
Cand{\`{e}}s, E.J.
\newblock {Compressive sampling}.
\newblock {In Proceedings of the International Congress of Mathematicians}, Madrid, Spain, 
  22--30 August 2006; pp. 1433--1452.

\bibitem[Candes \em{et~al.}(2008)Candes, Wakin, and Boyd]{Candes2008a}
Candes, E.J.; Wakin, M.B.; Boyd, S.P.
\newblock {Enhancing sparsity by reweighted L1 minimisation}.
\newblock {\em J. Fourier Anal. Appl.} {\bf 2008}, {\em
  14},~877--905.

\bibitem[Calabrese \em{et~al.}(2014)Calabrese, Badea, Coe, Lubach, Styner, and
  Johnson]{Calabrese2014}
Calabrese, E.; Badea, A.; Coe, C.L.; Lubach, G.R.; Styner, M.A.; Johnson, G.A.
\newblock {Investigating the tradeoffs between spatial resolution and diffusion
  sampling for brain mapping with diffusion tractography: Time well spent?}
\newblock {\em Hum. Brain Mapp.} {\bf 2014}, {\em 35},~5667--5685.

\bibitem[Zhan \em{et~al.}(2013)Zhan, Jahanshad, Ennis, Jin, Bernstein,
  Borowski, Jack, Toga, Leow, and Thompson]{Zhan2013}
Zhan, L.; Jahanshad, N.; Ennis, D.B.; Jin, Y.; Bernstein, M.A.; Borowski, B.J.;
  Jack, C.R.; Toga, A.W.; Leow, A.D.; Thompson, P.M.
\newblock {Angular Versus Spatial Resolution Trade-Offs for Diffusion Imaging
  Under Time Constraints}.
\newblock {\em Hum. Brain Mapp.} {\bf 2013}, {\em 32},~2688--2706.

\bibitem[Vos \em{et~al.}(2016)Vos, Aksoy, Han, Holdsworth, Maclaren, Viergever,
  Leemans, and Bammer]{Vos2016}
Vos, S.B.; Aksoy, M.; Han, Z.; Holdsworth, S.J.; Maclaren, J.; Viergever, M.A.;
  Leemans, A.; Bammer, R.
\newblock {NeuroImage Trade-off between angular and spatial resolutions in vivo
  fiber tractography}.
\newblock {\em NeuroImage} {\bf 2016}, {\em 129},~117--132.

\bibitem[Scherrer \em{et~al.}(2016)Scherrer, Afacan, Taquet, Prabhu, Gholipour,
  and Warfield]{Scherrer2016}
Scherrer, B.; Afacan, O.; Taquet, M.; Prabhu, S.P.; Gholipour, A.; Warfield, K.
\newblock {Super-resolution reconstruction to increase the spatial resolution
  of diffusion weighted images from orthogonal anisotropic acquisition}.
\newblock {\em HHS Pubblic Access} {\bf 2012}, 16, 1465--1476. 

\bibitem[Gao \em{et~al.}(2014)Gao, Li, Zhang, Zhou, and Hu]{Gao2014}
Gao, H.; Li, L.; Zhang, K.; Zhou, W.; Hu, X.
\newblock {PCLR: Phase-constrained low-rank model for compressive
  diffusion-weighted MRI}.
\newblock {\em Magn. Reson. Med.} {\bf 2014}, {\em
  72},~1330--1341.

\bibitem[Haldar \em{et~al.}(2013)Haldar, Wedeen, Nezamzadeh, Dai, Weiner,
  Schuff, and Liang]{Haldar2013}
Haldar, J.P.; Wedeen, V.J.; Nezamzadeh, M.; Dai, G.; Weiner, M.W.; Schuff, N.;
  Liang, Z.P.
\newblock {Improved diffusion imaging through SNR-enhancing joint
  reconstruction}.
\newblock {\em Magn. Reson. Med.} {\bf 2013}, {\em 69},~277--289.
\newblock
  doi:{\changeurlcolor{black}\href{https://doi.org/10.1002/mrm.24229}{\detokenize{10.1002/mrm.24229}}}.

\bibitem[Jeong \em{et~al.}(2003)Jeong, Kim, and Parker]{Jeong2003}
Jeong, E.K.; Kim, S.E.; Parker, D.L.
\newblock {High-resolution diffusion-weighted 3D MRI, using diffusion-weighted
  driven-equilibrium (DW-DE) and multishot segmented 3D-SSFP without navigator
  echoes}.
\newblock {\em Magn. Reson. Med.} {\bf 2003}, {\em 50},~821--829.
\newblock
  doi:{\changeurlcolor{black}\href{https://doi.org/10.1002/mrm.10593}{\detokenize{10.1002/mrm.10593}}}.

\bibitem[Cheng \em{et~al.}(2015)Cheng, Shen, Basser, and Yap]{Cheng2015}
Cheng, J.; Shen, D.; Basser, P.J.; Yap, P.T.
\newblock {Joint 6D k-q Space Compressed Sensing for Accelerated High Angular
  Resolution Diffusion MRI}.
\newblock  In \emph{International Conference on Information Processing in Medical Imaging}; Springer: Cham, Switzerland, 2015; Volume 9123, pp. 782--793.

\bibitem[M.~Mani and Jacob(2021)]{Magnotta2021}
Mani, M.M.M.; Jacob, M.
\newblock qModeL: A plug-and-play model-based reconstruction for highly
  accelerated multi-shot diffusion MRI using learned priors.
\newblock {\em Magn. Reson. Med.} {\bf 2021}, {\em 86},~835--851.

\bibitem[Ramos-Llord\'en \em{et~al.}(2020)Ramos-Llord\'en, Ning, C.~Liao,
  Michailovich, and K.~Setsompop]{Ramos2020}
Ramos-Llord\'en, G.; Ning, L.; Liao, R.M.C.; Michailovich, O.; Setsompop,
  Y.R.K.
\newblock High-fidelity, accelerated whole-brain submillimeter in vivo.
\newblock {\em Magn. Reson. Med.} {\bf 2020}, {\em 84},~1781--1795.

\bibitem[Sun \em{et~al.}(2015)Sun, Sakhaee, Entezari, and Vemuri]{Sun2015}
Sun, J.; Sakhaee, E.; Entezari, A.; Vemuri, B.C.
\newblock {Leveraging EAP-Sparsity for Compressed Sensing of MS-HARDI in
  (k,q)-Space}.
\newblock  In \emph{International Conference on Information Processing in Medical Imaging}; Springer: Cham, Switzerland, 2015; 
  Volume~24, pp. 375--386.

\bibitem[Awate and Dibella(2013)]{Awate2013}
Awate, S.P.; Dibella, V.R.E.
\newblock {Compressed sensing HARDI via rotation-invariant concise
  dictionaries, flexible k-space under-sampling, and multiscale spatial
  regularity}.
\newblock   In Proceedings of the 2013 IEEE 10th International Symposium on Biomedical Imaging, San Francisco, CA, USA,  7--11 April 2013; pp.
  9--12.
\newblock
  doi:{\changeurlcolor{black}\href{https://doi.org/10.1109/ISBI.2013.6556399}{\detokenize{10.1109/ISBI.2013.6556399}}}.

\bibitem[Combettes and Pesquet(2016)]{Combettes2016}
Combettes, P.L.; Pesquet, J.C.
\newblock {Stochastic forward-backward and primal-dual approximation algorithms
  with application to online image restoration}.
\newblock  In Proceedings of the 2016 24th European Signal Processing Conference (EUSIPCO),  Budapest, Hungary, \mbox{29 August--2 September 2016;}
  Volume~1, pp. 1813--1817.

\bibitem[{H. Gudbajartsson}(1995)]{H.Gudbajartsson1995}
{Gudbajartsson}, S.P.H. 
\newblock {The Rician Distribution of Noisy MRI Data}.
\newblock {\em Changes} {\bf 1995}, {\em 29},~997--1003.

\bibitem[Henkelman(1985)]{Henkelman1985}
Henkelman, R.M.
\newblock {Measurement of signal intensities in the presence of noise in MR
  images}.
\newblock {\em Med. Phys.} {\bf 1985}, {\em 12},~232--233.

\bibitem[Chen \em{et~al.}(2001)Chen, Donoho, and Saunders]{Chen2001}
Chen, S.S.; Donoho, D.L.; Saunders, M.A.
\newblock {Atomic Decomposition by Basis Pursuit}.
\newblock {\em SIAM J. Sci. Comput.} {\bf 2001}, {\em
  43},~129--159.

\bibitem[Donoho(1995)]{Donoho1995}
Donoho, D.L.
\newblock {De-Noising by Soft-Thresholding}.
\newblock {\em IEEE Trans. Inf. Theory} {\bf 1995}, {\em
  41},~613--627.
\newblock
  doi:{\changeurlcolor{black}\href{https://doi.org/10.1109/18.382009}{\detokenize{10.1109/18.382009}}}.

\bibitem[Ochs \em{et~al.}(2015)Ochs, Dosovitskiy, Brox, and Pock]{Ochs2015}
Ochs, P.; Dosovitskiy, A.; Brox, T.; Pock, T.
\newblock On iteratively reweighted algorithms for non-smooth nonconvex
  optimization in computer vision. \emph{SIAM J. Imaging Sci.} {\bf 2015},
\newblock {\em 8},~331--372.

\bibitem[Geiping and Moeller(2018)]{Geiping2018}
Geiping, J.; Moeller, M.
\newblock Composite optimization by nonconvex majorization-minimization. \emph{SIAM J. Imaging Sci.} {\bf
  2018},
\newblock {\em 11},~2494--2598.

\bibitem[Ochs \em{et~al.}(2019)Ochs, Fadili, and Brox]{Ochs2019}
Ochs, P.; Fadili, J.; Brox, T.
\newblock Non-smooth non-convex {B}regman minimization: {U}nification and new
  algorithms. \emph{J. Optim. Theory Appl.} {\bf 2019},
\newblock {\em 181},~244--278.

\bibitem[Repetti and Wiaux(2019)]{Repetti2019b}
Repetti, A.; Wiaux, Y.
\newblock Variable Metric Forward-Backward Algorithm for Composite Minimization
  Problems.
\newblock \emph{SIAM J. Optim.} \textbf{2021}, \emph{31}, 1215--1241.

\bibitem[Combettes and Wajs(2005)]{Combettes2005}
Combettes, P.L.; Wajs, V.R.
\newblock {Signal Recovery by Proximal Forward-Backward Splitting}.
\newblock {\em Multiscale Model. Simul.} {\bf 2005}, {\em
  4},~1168--1200.

\bibitem[Caruyer \em{et~al.}(2013)Caruyer, Lenglet, Sapiro, and
  Deriche]{Caruyer2013}
Caruyer, E.; Lenglet, C.; Sapiro, G.; Deriche, R.
\newblock Design of multishell sampling schemes with uniform coverage in
  diffusion MRI.
\newblock {\em Magn. Reson. Med.} {\bf 2013}, {\em
  69},~1534--1540,
\newblock
  doi:10.1002/mrm.24736.

\bibitem[Miller and Pauly(2003)]{Miller2003}
Miller, K.L.; Pauly, J.M.
\newblock {Nonlinear phase correction for navigated diffusion imaging}.
\newblock {\em Magn. Reson. Med.} {\bf 2003}, {\em 50},~343--353.

\bibitem[Pipe \em{et~al.}(2002)Pipe, Farthing, and Forbes]{Pipe2002}
Pipe, J.G.; Farthing, V.G.; Forbes, K.P.
\newblock {Multishot diffusion-weighted FSE using PROPELLER MRI}.
\newblock {\em Magn. Reson. Med.} {\bf 2002}, {\em 47},~42--52.

\bibitem[Roemer \em{et~al.}(1990)Roemer, Edelstein, Hayes, Souza, and {Muller.
  O. M.}]{Roemer1990}
Roemer, P.B.; Edelstein, W.A.; Hayes, C.E.; Souza, S.P.; {Muller, O.M.}
\newblock {The NMR Phase Array}.
\newblock {\em Magn. Reson. Med.} {\bf 1990}, {\em 16},~192--225.


\bibitem[Hein \em{et~al.}(2016)Hein, Neher, Christophe, and
  Alexandre]{Hein2016}
Maier-Hein, K.H.; Neher, P.F.; Houde, J-C.; C{\^{o}}t{\'{e}}, M-A.; Garyfallidis, E.; Zhong, J.; Chamberland, M.; Yeh, F-C.; Lin, Y-C.; Ji, Q.; Reddick, W.E.; Glass, J.O.; Chen, D.Q.; Feng, Y.; Gao, C.; Wu, Y.; Ma, J.; He, R.; Li, Q.; Westin, C-F.; Deslauriers-Gauthier, S.; Gonz{\'{a}}lez, J O-O.; Paquette, M.; St-Jean, S.; Girard, G.; Rheault, F.; Sidhu, J.; Tax, C.; Guo, F.; Mesri, H-Y.; D{\'{a}}vid, S.; Froeling, M.; Heemskerk, A-M.; Leemans, A.; Bor{\'{e}}, A.; Pinsard, B.; Bedetti, C.; Desrosiers, M.; Brambati, S.; Doyon, J.; Sarica, A.; Vasta, R.; Cerasa, A.; Quattrone, A.; Yeatman, J.; Khan, A-R.; Hodges, W.; Alexander, S.; Romascano, D.; Barakovic, M.; Aur{\'{i}}a, A.; Esteban, O.; Lemkaddem, A.; Thiran, J-P.; Cetingul, H-E.; Odry, B-L.; Mailhe, B.; Nadar, M-S.; Pizzagalli, F.; Prasad, G.; Villalon-Reina, J-E.; Galvis, J.; Thompson, P-M.; Requejo, F-D-S.; Laguna, P-L.; Lacerda, L-M.; Barrett, R.; Dell'Acqua, F.; Catani, M.; Petit, L.; Caruyer, E.; Daducci, A.; Dyrby, T-B.; Holland-Letz, T.; Hilgetag, C-C.; Stieltjes, B.; Descoteaux, M.
\newblock {The challenge of mapping the human connectome based on diffusion tractography}.
\newblock {\em Nature Communications} {\bf 2017}, 8, 1--23.



\bibitem[Komodakis and Pesquet(2014)]{Komodakis2014}
Komodakis, N.; Pesquet, J.C.
\newblock {Playing with Duality: An Overview of Recent Primal-Dual Approaches
  for Solving Large-Scale Optimization Problems}. \emph{IEEE Signal Process. Mag.} \textbf{2015}, \emph{32}, 31--54.

\bibitem[Beck and Teboulle(2009)]{Beck2009}
Beck, A.; Teboulle, M.
\newblock {A Fast Iterative Shrinkage-Thresholding Algorithm for Linear Inverse
  Problems}.
\newblock {\em SIAM J. Imaging Sci.} {\bf 2009}, {\em 2},~183--202.

\end{thebibliography}
\end{document}